\RequirePackage{fix-cm}
\documentclass[smallcondensed]{svjour3}     
\smartqed  
\usepackage[T1]{fontenc}
\usepackage[utf8]{inputenc}
\usepackage[english]{babel}
\usepackage{amssymb}
\usepackage{amsmath}
\usepackage{latexsym}
\usepackage{mathtools}
\usepackage{graphicx}
\usepackage{bm}
\usepackage{amsbsy}
\usepackage{wasysym}
\usepackage{booktabs}
\usepackage{subcaption}
\usepackage{caption}
\usepackage{tabularx}
\usepackage{epsfig}
\usepackage{natbib}
\usepackage[pagewise]{lineno}
\usepackage{placeins}
\usepackage{hyperref}
\hypersetup{
    colorlinks=true,
    linkcolor=blue,
    filecolor=blue,      
    urlcolor=blue,
    citecolor=blue,
}
\usepackage{upgreek}
\journalname{Experimental Astronomy}

\graphicspath{{Figures/}} 

\def\arxivprefixesep{:}

\newcommand{\eprint}[2][]{%
{\tt\if!#1!#2\else#1\arxivprefixesep\ignorespaces#2\fi}%
}

\newcommand{\Msun}{\hbox{$\mathrm{M}_{\rm{\odot}}$}}

\newcommand{\orcid}[1]{\href{https://orcid.org/#1}{\includegraphics[width=8pt]{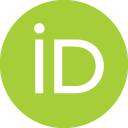}}}

\begin{document}

\title{Spectral performance of the Microchannel X-ray Telescope on board the SVOM mission}
\titlerunning{Spectral performance of the MXT on board the SVOM mission} 

\author{B.~Schneider\textsuperscript{*1}\orcid{0000-0003-4876-7756} \and
        N.~Renault-Tinacci\textsuperscript{1}\orcid{0000-0001-8443-8007} \and
        D.~Götz\textsuperscript{1} \and
        A.~Meuris\textsuperscript{1} \and
        P.~Ferrando\textsuperscript{1} \and
        V.~Burwitz\textsuperscript{2} \and
        E.~Doumayrou\textsuperscript{1} \and
        T.~Lavanant\textsuperscript{1} \and
        N.~Meidinger\textsuperscript{2} \and
        K.~Mercier\textsuperscript{3}
}
\authorrunning{B.~Schneider} 

\institute{
\textsuperscript{1}Universit\'e Paris-Saclay, Université Paris Cit\'e, CEA, CNRS, AIM, 91191, Gif-sur-Yvette, France \\
\textsuperscript{2}Max-Planck-Institut f\"{u}r Extraterrestrische Physik, 85748 Garching, Germany \\
\textsuperscript{3}Centre National d’Etudes Spatiales, Centre Spatial de Toulouse, Toulouse Cedex 9, France \\
\textsuperscript{*} \email{benjamin.schneider@cea.fr}
}

\date{Received: XXXX / Accepted: YYYY}

\maketitle

\begin{abstract}
    The Microchannel X-ray Telescope (MXT) is an innovative compact \hbox{X-ray} instrument on board the SVOM astronomical mission dedicated to the study of transient phenomena such as gamma-ray bursts. During 3 weeks, we have tested the MXT flight model at the Panter X-ray test facility under the nominal temperature and vacuum conditions that MXT will undergo in-flight. 
    We collected data at series of characteristic energies probing the entire MXT energy range, from 0.28~keV up to 9~keV, for multiple source positions with the center of the point spread function (PSF) inside and outside the detector field of view (FOV). 
    We stacked the data of the positions with the PSF outside the FOV to obtain a uniformly illuminated matrix and reduced all data sets using a dedicated pipeline. We determined the best spectral performance of MXT using an optimized data processing, especially for the energy calibration and the charge sharing effect induced by the pixel low energy thresholding. Our results demonstrate that MXT is compliant with the instrument requirement regarding the energy resolution (\hbox{$<$80~eV} at 1.5~keV), the low and high energy threshold, and the accuracy of the energy calibration ($\pm$20~eV). We also determined the charge transfer inefficiency ($\sim$$10^{-5}$) of the detector and modeled its evolution with energy prior to the irradiation that MXT will undergo during its in-orbit lifetime. Finally, we measured the relation of the energy resolution as function of the photon energy. We determined an equivalent noise charge of $4.9 \pm 0.2 \ \mathrm{e}^{-}_{\mathrm{rms}}$ for the MXT detection chain and a Fano factor of $0.131 \pm 0.003$ in silicon at 208 K, in agreement with previous works.
    This campaign confirmed the promising scientific performance that MXT will be able to deliver during the mission lifetime.
\end{abstract}

\section{Introduction}
\label{sec:Introduction}
    Gamma-ray bursts (GRBs) are the most luminous and energetic events in the Universe, produced by the collapsing of high-mass stars ($>$50~\Msun) or by the coalescence of a binary system of compact objects (e.g., the merger of two neutron stars), see \cite{levan2016a} for a review. In a few seconds, they are able to release a large amount of high-energy radiation, including X-ray and $\gamma$-ray photons, with typical isotropic equivalent luminosities of $L_{\mathrm{iso}} \sim 10^{50}-10^{53} \  \mathrm{erg\,s}^{-1}$. Their short-lived and unpredictable nature makes them challenging to detect and require the use of dedicated space-based instruments. The observation strategy used by the \textit{Swift} \citep{gehrels2004a} satellite launched in 2004 has deeply revolutionized the way of studying GRBs and has permitted to detect and localize more than 1\,000 GRBs. The Sino-French SVOM \citep[Space based multi-band astronomical Variable Objects Monitor,][]{wei2016a,atteia2022a} mission is the upcoming successor to \textit{Swift} dedicated to the study of GRBs and high-energy transients with an expected launch in 2023. The first scientific objective of SVOM is to detect, localize and perform multi-wavelength observations of GRBs. With its anti-solar pointing strategy and synergy between space and ground-based instruments, it is expected to provide a more complete and unbiased sample of well-characterized GRBs, especially concerning the GRB redshift measurements (with a goal of 2/3 of the full sample). SVOM carries a total of four instruments including two wide field of view (FOV) high-energy instruments, ECLAIRs and the Gamma Ray-burst Monitor (GRM), and two narrow field instruments, the Microchannel X-ray Telescope (MXT) and the Visible Telescope (VT). The GRB prompt emission is initially detected, localized and characterized by ECLAIRs, thanks to its large FOV of $\sim $2~sr and a broad energy range ($4\text{--}150$~keV). The satellite is then slewed towards the direction of the ECLAIRs error box, and the MXT and VT are used to improve the GRB localization and further characterize its emission. \\
    The MXT is a compact and light X-ray telescope focusing photons in the $0.2\text{--}10$~keV energy band. It will be able to detect and localize (within a few tens of arc seconds) the majority of GRBs, including those with faint or no optical afterglow.   
    After years of development and prototype models, the MXT has reached its final stage with the delivery of its flight model (FM). To evaluate and validate the imaging and spectral performance of the FM instrument prior to launch and in-orbit operations, we performed an end-to-end test campaign at the Panter X-ray test facility\footnote{\url{https://www.mpe.mpg.de/heg/panter}} \citep{burwitz2013a,bradshaw2019a} of the MPE, located in Neuried on the southwest part of Munich (Germany). During the campaign, we fully characterized and evaluated the performance of the instrument for multiple camera and optics temperatures, beam energies and point spread function (PSF) positions. In this paper, we describe the data processing algorithms used to reduce the data and improve the calibration process, and we present the spectral performance of the MXT FM camera.   \\
    The paper is organized as follows. In Section~\ref{sec:the_microchannel_x_ray_telescope_on_board_svom}, we describe the MXT design, the readout sequence and the on board data processing. Section~\ref{sec:experimental_setup_and_analysis_method} presents the Panter facility and analysis methods used to determine the spectral performance. We describe in Section~\ref{sec:experimental_results_and_spectral_performance} the results of the campaign for the electronic noise, the energy resolution and the charge transfer efficiency. Finally the conclusions are presented in Section~\ref{sec:conclusions}.

\section{The Microchannel X-ray Telescope on board SVOM} 
\label{sec:the_microchannel_x_ray_telescope_on_board_svom}
    \subsection{Instrument design overview}
    \label{sub:instrument_design_overview}
    The MXT is an innovative instrumental concept of X-ray optic based on micropore optics (MPOs) arranged in lobster-eye configuration \citep{fraser2010a,gotz2014a,feldman2017a}. This focusing technique allows, with a light and compact instrument (42~kg), to reach the sensitivity required to localize $\sim $80\% of the GRBs within a 2~arcmin precision after 10~minutes of observation. The MXT instrument characteristics derived from data collected during the Panter campaign and the expected scientific performance obtained from simulations are summarized in the Table~\ref{tab:mxt_performance}. Further detailed on the data analysis and the simulation processes are provided in the accompanying papers of \cite{gotz2022a} and Feldman et al. (2022). MXT will also provide accurate time-resolved spectral diagnostics of the X-ray emissions of transient sources (e.g., GRBs) or compact objects thanks to a focal plane at the state-of-the-art of X-ray space astronomy. The telescope, placed in the platform interface module, is composed of an optics module (MOP), a camera (MCAM), a telescope tube in carbon fiber, and a radiator. Two data processing units (MDPUs), in cold redundancy, are connected to the camera. The camera includes a focal plane assembly, a calibration wheel assembly, a front-end electronics assembly and a mechanical support structure. To cool down the detector and maintain its temperature at its operational value of $-65^{\circ}$C along the low-Earth orbit of SVOM, the focal plane includes an active cooling system based on three thermoelectric coolers connected through propylene heat pipes to the MXT radiator. The wheel assembly has three main positions: (i) a calibration position with a radioactive $^{55}$Fe source fully illuminating the detector, (ii) an open position for sky observations, and (iii) a closed position with a 10~mm thick copper shutter to ensure the protection of the detector against radiation damage during the regular passages through the South-Atlantic Anomaly.
    \begin{table}[b]
    \centering
    \caption{MXT instrument characteristics.}
    \renewcommand{\arraystretch}{1.25}
        \begin{tabular}{ll}
            \hline \hline
            Energy range             & $0.2 $--$ 10$~keV \\
            Field of View            & $58 \times 58$~arcmin \\
            Angular resolution       & 10~arcmin at 1.5~keV \\
            Source location accuracy & $<$120~arcsec for 80\% GRBs \\
            Effective area           & $\sim $35 cm$^2$ at 1.5~keV \\
            Sensitivity ($5 \sigma$) & 10~mCrab in 10~s \\
                                     & 150~$\upmu$Crab in 10~ks \\
            Energy resolution        & $<$80 eV at 1.5~keV \\
            Time resolution          & 100~ms \\
            \hline
        \end{tabular}
    \label{tab:mxt_performance}
    \end{table}
    
    \subsection{Focal plane readout and on board data processing}
    \label{sub:on_board_data_pre-processing}
\begin{figure}[t]
        \centering
        \includegraphics[width=\hsize]{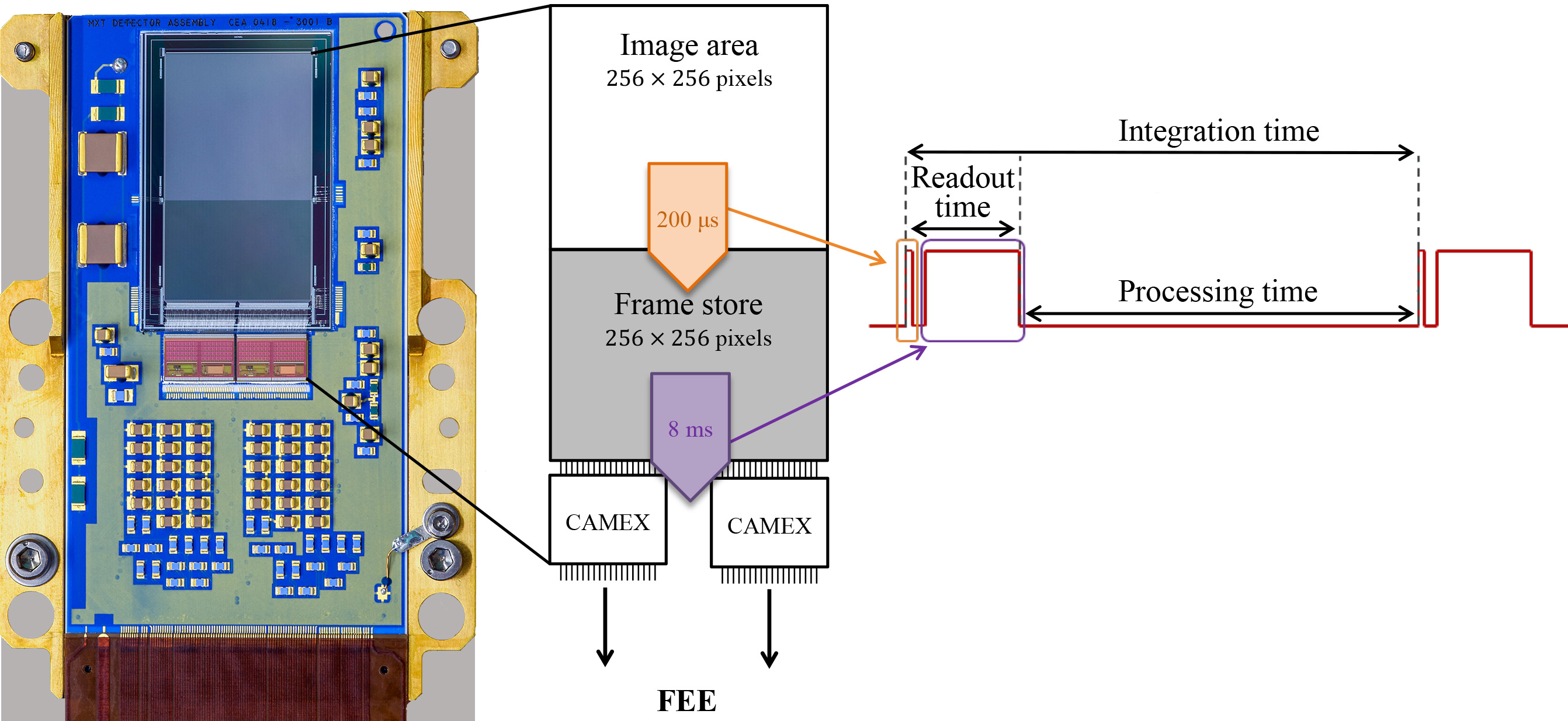}
        \caption{Front side picture of the MXT detector assembly. The pnCCD is composed of two areas, image and frame-store, which differ by their color shade in the picture. The two CAMEX ASICs are visible in the bottom part of the frame-store area. On the right, a schematic of the detector with the typical sequence to collect a frame and process it in the FEE is shown.}
        \label{fig:MXT_detector}
    \end{figure}

    The focal plane is based on a $256 \times 256$ pixels pnCCD and is read out by two 128-channels CAMEX ASICs, both provided by the Max Planck Institute for Extraterrestrial Physics (MPE) (see left panel of Fig.~\ref{fig:MXT_detector}).
    The design of the pnCCD is an upgrade of the pnCCD of XMM-Newton and includes an unexposed frame store area to reduce out-of-time events, as implemented in the eROSITA \citep{meidinger2010a} instrument on board the Spectrum-Roentgen-Gamma (SRG) mission. The ASIC is a direct heritage of the eROSITA instrument. The front-end electronics (FEE) provides all bias voltages for the CAMEX ASIC and the pnCCD as well as the readout control signals. The FEE also ensures the amplification and digitization of the analog channels, and performs a mode-dependent processing of the pixel data before transmission to the MDPU. The image area integrates photon events converted into charges during 100~ms. Then charges are rapidly transferred into a frame store area in 200~$\upmu$s. The frame store is read out row by row by transferring the charges to the anodes of the pnCCD columns. The signals of each column are converted into voltages by the on-chip JFETs and then amplified and filtered by the CAMEX analog channels; they are then timely multiplexed at the CAMEX analog outputs during the processing of the next row. This allows a readout of the frame store in 8~ms. The frame rate of 10 images per second has been chosen as a good compromise between the photon incident rate and the duty cycle of the CAMEX to reduce heating power in the focal plane, the main (analog) stages of the CAMEX being switch-off during integration and turned on just for the read-out phase. \\
    A frame corresponds to 1~Mbits data and is at the limit of the capacity of the SpaceWire datalink between the MCAM and the MDPU. As a consequence, a real-time processing is implemented on the frame data to extract and transmit only the ‘events’, defined as the pixels containing a significant signal. This ‘event mode’ is the nominal readout mode and is performed at a speed of 10 frames per second. This processing is done by the FEE at the pixel level and consists in: (i) subtracting a pixel dependent offset value to its raw amplitude, (ii) subtracting a frame dependent common-mode noise value (calculated for each CAMEX row of each frame), and (iii) comparing the resulting amplitude to a pixel dependent low-level threshold (LLT) set to $k$ times the noise level value of the pixel, with $k$ being programmable. If the pixel signal is above the threshold, the pixel is considered as a true signal, produced by an X-ray or an ionizing particle, and not as a noise fluctuation. 
    Practically, $k = 4$ was determined to be a good compromise to ensure the extraction of X-ray events and obtain the minimal energy resolution. \\
    The transmission of all data collected by the exposed area (image area) to the MDPU is also possible in a ‘full frame mode’ with a decimation factor of two (5~images per second but maintaining the integration time of 100~ms in the detector). This mode is used on board to perform ‘dark’ measurements with the wheel in close position and from which the MDPU calculates the offsets and LLT values for each pixel. The two resulting tables are then uploaded to the FEE for the ‘event mode’ processing discussed above.

\section{Experimental setup and analysis method}
\label{sec:experimental_setup_and_analysis_method}
    \subsection{Panter calibration campaign}
    \label{sub:panter_calibration_campaign}
    \begin{figure}[t]
        \begin{subfigure}[c]{0.5\hsize}
            \centering
            \includegraphics[width=0.8\hsize]{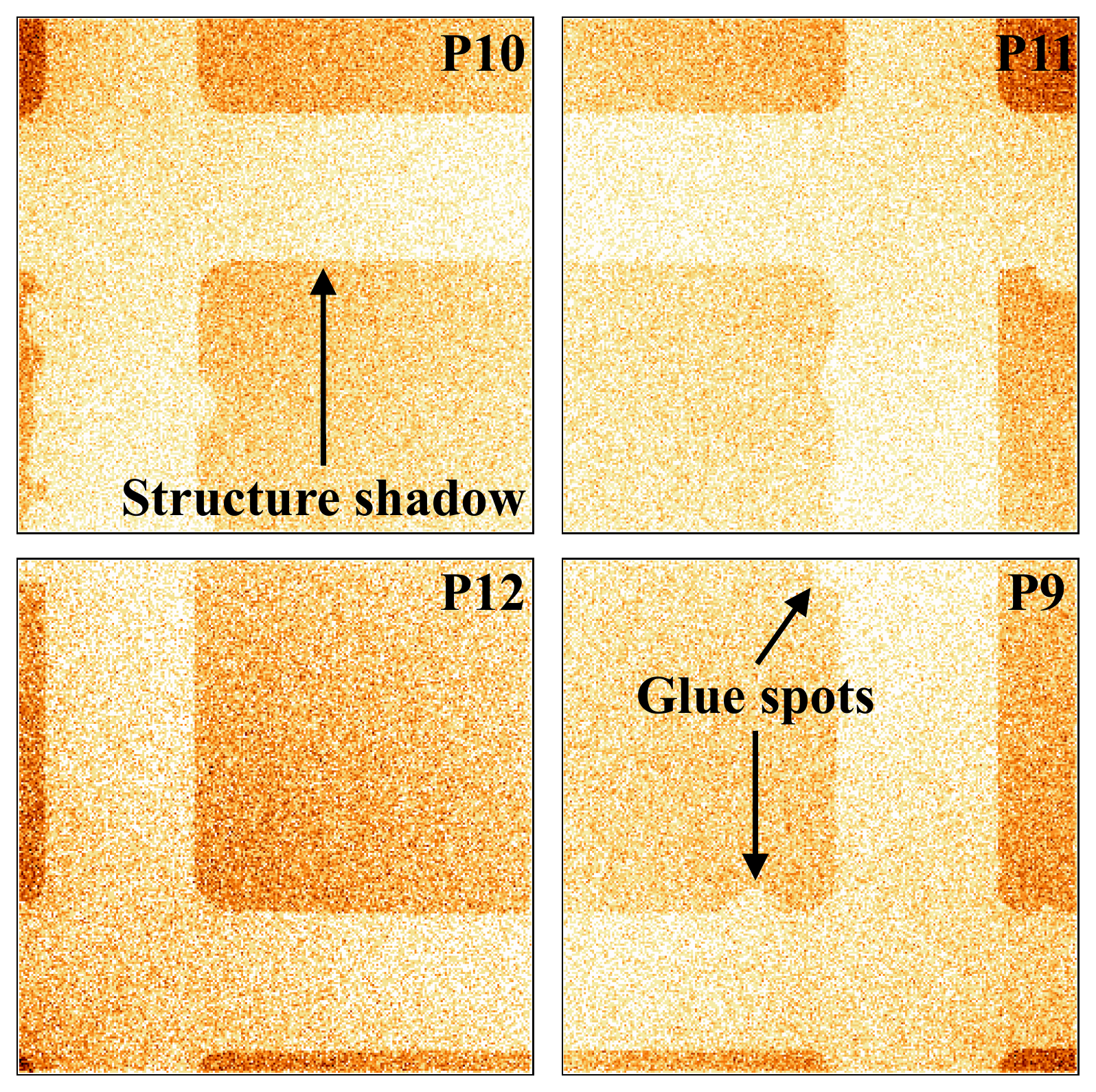}
            \vspace*{0.31cm}
        \end{subfigure}
        \hfill
        \begin{subfigure}[c]{0.5\hsize}
            \centering
            \includegraphics[width=\hsize]{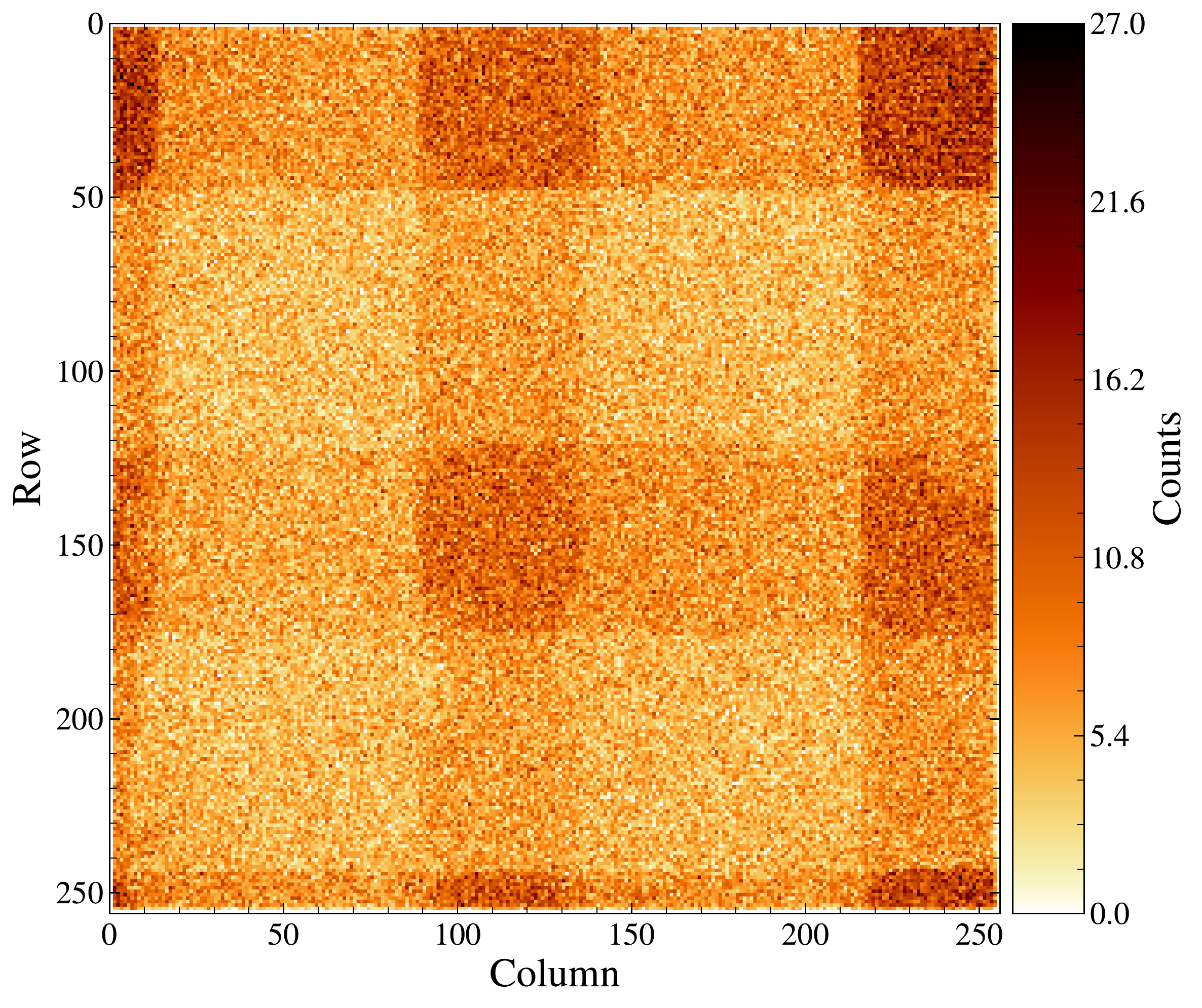}
        \end{subfigure}
        \caption{Left panel: Count maps of the four PSF positions (P9-P12) outside the FOV detector at 1.49~keV (Al-K) and a detector temperature of $-65^{\circ}$C. The lighter vertical and horizontal area (lower counts number) represents the shadow of the mechanical structure holding the MPO. Along the shadows, we can moreover identify small spots produced by the glue used during the optics manufacturing process.
                 Right panel: Count map resulting from the stacking of the count maps P9, P10, P11 and P12 visible on the left panel.}
        \label{fig:countmap}
    \end{figure}
    From October 20\textsuperscript{th} to November 5\textsuperscript{th} 2021, the MXT FM instrument has been intensively tested and characterized at the Panter X-ray test facility. 
    It consists of an X-ray tube producing energy lines from 0.28 to 10~keV and a vacuum chamber for the focal plane instrumentation, separated from each other by a 130-m-long and 1-m-diameter vacuum beamline.
    This configuration allows to produce an almost point-like source (i.e., a quasi-parallel beam), and the chamber provides the environmental space conditions of pressure and temperature at which the MXT will be operated in flight. \\
    The instrument was in its final flight configuration, including the FM Optics, the FM Camera and the nominal and redundant MDPUs. We acquired data for nine source positions (P0-P8) inside the detector FOV, separated by 15 to 50~arcmin from the center, and four source positions (P9-P12) with the PSF centre outside the detector FOV (i.e., with the X-ray flux passing through the MPOs without being reflected) to obtain a more uniform illumination of the detector. The left panel of the Fig.~\ref{fig:countmap} shows for an energy of 1.49~keV (Al-K) the count maps of the four corner positions outside the FOV. The count maps show a good uniform illumination overall, except in some areas where we can clearly see the shadow of the mechanical structure holding the MPOs, which absorbs a significant fraction of the X-ray flux. \\
    CCD detectors such as the one used in MXT, accumulate photons in the image area during a fixed ‘integration time’ (100~ms in MXT ‘event mode’ configuration, see Section~\ref{sub:on_board_data_pre-processing}). Then, the charges created by X-ray photons are quickly transferred to the frame store (200~$\upmu$s) and read out by the electronics (10~ms). During the integration time, several photons can hit the same pixels and produce a quantity of charge equal to the sum of the photons, a phenomenon called pile-up. The X-ray spectra generated by the Panter facility are not purely monochromatic and \textit{Bremsstrahlung} and by-products created by the X-ray tube may be present in addition to the selected energy. In a pile-up regime with multiple energy lines, it becomes complex and difficult to disentangle the contributions of each photon and thus to reconstruct the individual photon energy. We minimized this phenomenon by tuning the flux of the X-ray source before taking our calibration data. Because the MXT effective area is energy-dependent, the flux optimization was performed for each selected energy (Table~\ref{tab:energy_lines}). We tuned the photon counting rate to have less than 0.01 count/pixel/frame in the central PSF core for positions inside the FOV and in the entire frame for positions outside the FOV. This ensures to minimize the pile-up effect \citep[$<$1\%, see][]{ballet1999a} and preserves a reasonable acquisition time (a few minutes) to reach a statistic of approximately $20\,000$ and $64\,000$ photons for P\,<\,P9 and P\,$\ge$\,P9, respectively. \\
    We derived and characterized the detector response over the entire MXT energy range by stacking the data from positions P9 to P12 (as shown in the right panel of Fig.~\ref{fig:countmap} for Al-K). The resulting data sets allow us to have a sufficient uniform coverage of the detector. We collected between 1\,500 and 3\,000 counts per column (i.e., more than 284\,000 photons over the entire detector), exceeding the minimum of 1\,000 counts per columnn required for the energy calibration. At the Cu-K line energy, the instrument effective area becomes too small for keeping a reasonable acquisition time with the out-of-FOV positions, and we thus used the sum of the in-FOV positions from P0-P8. The resulting count map is not as uniform as for the lower energies, due to the presence of the PSF core inside the FOV of the detector, but the minimum of 1\,000 counts per column criterion is satisfied.

    \subsection{Energy calibration method}
    \label{sub:Energy calibration method}
        \subsubsection{Event extraction} 
        \label{ssub:event_extraction}
        When an X-ray photon interacts with the detector, it induces via photoelectric effect an amount of charge at its interaction position that is proportional to its energy. Depending on the photon hit position, the collected charges at the electrodes can be spread over up to $2 \times 2$ pixels and generate a pattern. In Section~\ref{sub:on_board_data_pre-processing}, we describe the on board pre-processing performed by the FEE and MDPU. The resulting data consists of a list of hit pixels with a deposited energy above the defined threshold (i.e., LLT). Our core analysis algorithms are mainly based on those developed by the MPE for the eROSITA detector and described by \cite{andritschke2008a}, but see also \cite{ceraudo2020a} for more details on the MXT analysis algorithms. First, the neighboring pixels are grouped from the events list by using an improved and optimized \texttt{Python} routine based on the \textit{scipy.ndimage} package. The design of the MXT detector (pixel size, detector thickness, voltage) implies that collected electrons can be shared from 1 to 4 adjacent pixels, leading to a number of 13 unique valid patterns (X-ray event). Group of pixels are then classified with a pattern code following the standard convention previously used by the XMM/EPIC and other X-ray space instruments. It may happen that the energy of the incoming photon is not fully contained in the hit pixels returned by the FEE. This occurs if the charge sharing between pixels leads to a signal below the FEE low-level threshold for the adjacent pixel(s) to the main charge distribution peak. This problem and its implications on the spectral performances are further discussed in Section~\ref{sub:management_of_multiple_events}.

        \subsubsection{Calibration method} 
        \label{ssub:calibration_method}
        
        \begin{figure}
        \begin{minipage}[b]{0.69\textwidth}
            \centering
            \includegraphics[width=0.95\hsize]{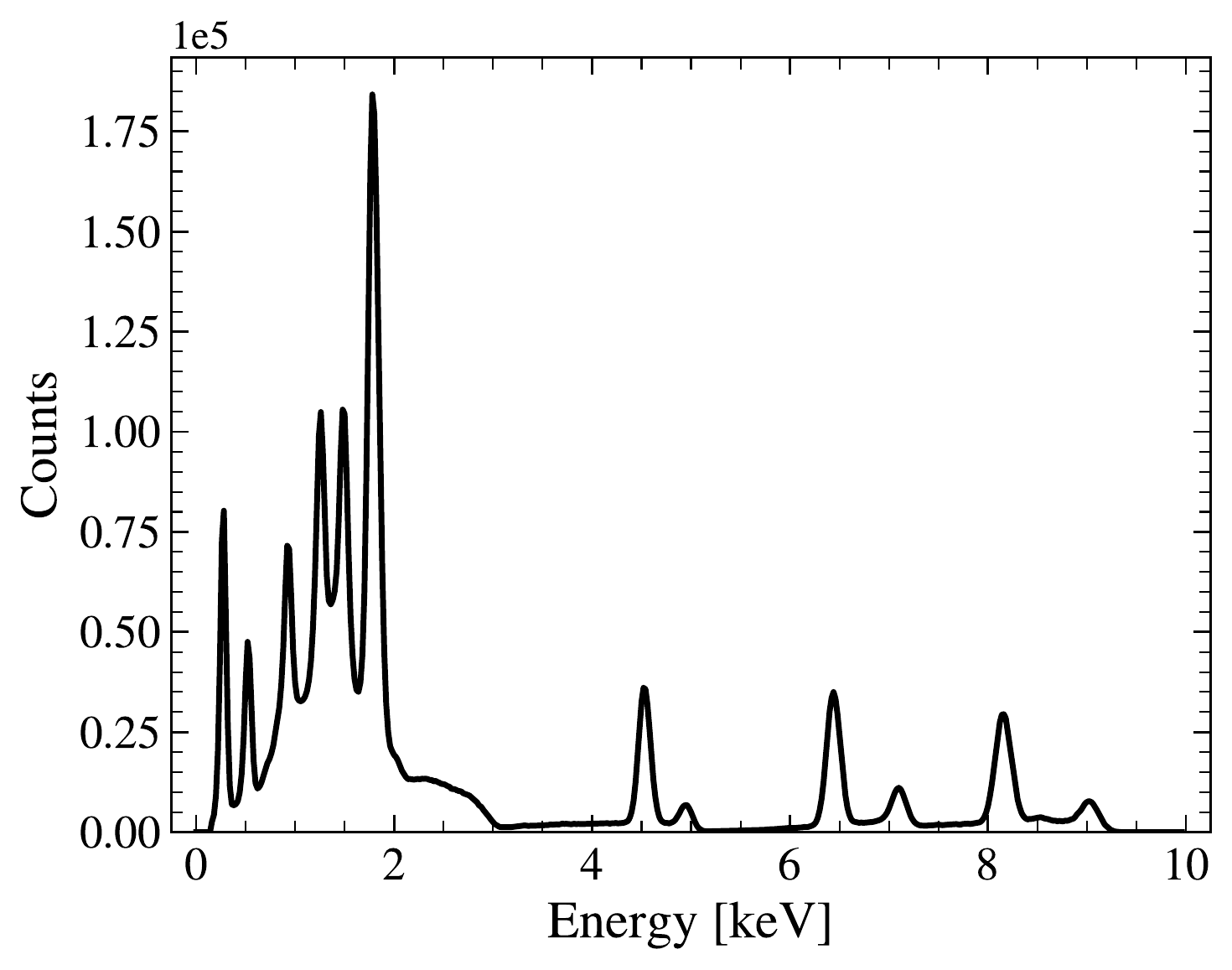}
            \caption{Combined spectrum used to derive the energy calibration. Only well-resolved fluorescence lines produced by the Panter X-ray source (Table~\ref{tab:energy_lines}) are considered.}
            \label{fig:combined_spectrum}
        \end{minipage}
        \hfill
        \begin{minipage}[b]{0.28\textwidth}
            \centering            
            \captionsetup{type=table} 
            \renewcommand{\arraystretch}{1.25}
            \begin{tabular}{ll}
                \hline \hline
                Element & Energy [eV] \\
                \hline 
                C-K        & 277 \\
                O-K        & 525 \\
                Cu-L       & 930 \\
                Mg-K       & 1253 \\
                Al-K       & 1486 \\
                W-M        & 1774 \\ 
                Ti-K$\alpha$     & 4510 \\
                Ti-K$\beta$      & 4950 \\
                Fe-K$\alpha$     & 6400 \\
                Fe-K$\beta$      & 7053 \\
                Cu-K$\alpha$     & 8040 \\
                Cu-K$\beta$     & 8910 \\
                \hline 
            \end{tabular}
            \caption{Fluorescence lines produced by the Panter X-ray source and used for the spectral calibration. Energies are extracted from the X-Ray Data Booklet \citep{thompson2009a}.}
            \label{tab:energy_lines}
        \end{minipage}
        \end{figure}
        The energy calibration consists in finding a relation between the digitized detector signal in Analog-to-Digital Units (ADU) and the incoming energy photons in keV. This process requires prior knowledge on the X-ray source energy irradiating the detector (e.g., a radioactive isotope, an X-ray tube, or a well-studied astrophysical source). A common calibration method is the \textit{peak fitting} approach which consists in finding the energy lines of the un-calibrated spectrum and fitting their peak center with a Gaussian profile. The calibration parameters are then obtained by determining the relation between the known theoretical peak positions (in keV) and the derived peak positions (in ADU). 
        We opted for an alternative calibration method, which is significantly faster and more efficient in low count regimes or low X-ray energy lines.
        We calibrated our data sets with the \textit{Energy calibration via correlation} (ECC) method, introduced by \cite{maier2016a} and more particularly its fastest upgraded version using adaptive mesh refinement (AMR) to discretize the parameter space \citep{maier2020a}. The method is based on the maximum of correlation found between a synthetic spectrum of the reference source and the un-calibrated observed spectrum. As mentioned in Section~\ref{sub:on_board_data_pre-processing}, two CAMEX ASICs with 128 channels each read out the detector. This implies that the calibration parameters (i.e., gains and offsets) depend mainly on the column. To avoid mixing data with different gains (column dependent), we performed the calibration by using only the single events (the doubles along one column might also be used to increase the statistic). As mentioned above, we stacked the data from positions P9 to P12 for each Panter energy line to have a matrix illuminated as homogeneously as possible. In addition, to improve the calibration and get the optimal solutions, we combined in a single spectrum all well-resolved lines available over the entire energy range of the detector (see Table~\ref{tab:energy_lines}). The combined spectrum used to determine the calibration parameters is shown in Fig.~\ref{fig:combined_spectrum}. 
        The synthetic spectrum was constructed using energy lines of Table~\ref{tab:energy_lines} and their relative intensities were defined as the ones observed in the combined spectrum in ADU unit. Then, each line was convolved with a Gaussian function adapted to the expected spectral resolution of MXT. No specific background was added to match the observed \textit{Bremsstrahlung} in the calibration spectrum.

        \subsubsection{Charge transfer (in-)efficiency correction} 
        \label{ssub:charge_transfer_efficiency_correction}
        When the electrons collected in the image area are progressively moved and transferred row-by-row to the anode, a fraction of the charge packets might be captured by traps from crystal defects. This phenomenon named charge transfer efficiency (CTE) or equivalently charge transfer inefficiency ($\text{CTI} = 1 - \text{CTE}$) is cumulative at each transfer and therefore the most distant row from the anode is the most affected. 
        The energy of the reconstructed photons is thus slightly underestimated and the center of the lines are shifted to a lower energy. In addition, the spectral performance (e.g., the energy resolution) might be degraded. Even if this effect is negligible at first order, it is expected to increase with time due to the radiation in space impacting the detector.
        Theoretical calculation of the CTE is complex and challenging because it depends on many parameters, such as operating conditions (temperature, operating voltage), detector defects (material, radiation damage) or X-ray source properties (energy, photon flux), which makes the empirical approach more appropriate.
        We evaluated the initial CTE of the detector at a given energy by using similar data sets to the ones used in Section~\ref{ssub:calibration_method} (data from P9 to P12). Once the data sets are calibrated in energy, we determined the line position centers for the 256-row spectra constructed from single events (double along row might also be used to increase the statistic). Finally, we derived the CTE by fitting the line centroids as a function of the number of transfers using the model described by \cite{ceraudo2020a}.

    \subsection{Management of multiple events}
    \label{sub:management_of_multiple_events}
        If charges produced by an X-ray photon extend over several pixels, the photon energy is reconstructed during the post-data processing by summing the individual energy in pixels. When one or several pixels have a lower energy than the LLT value (i.e., the threshold for suppressing noise events), a fraction of the photon energy gets lost and the recombined photon energy is thus slightly underestimated. This produces a charge sharing (CS) energy effect, which on one side induces a shift of the line position to lower energy, and on the other side degrades the spectral performance of the instrument. For a given energy, the fraction of energy loss depends on the event multiplicity and affects therefore particularly the spectral performance of the spectrum combining all event types. The allocation of the incident photon energy to each pixel involved in a multiple event was introduced by \cite{dennerl2012a} to improve the spatial resolution of the eROSITA instrument.   
        We estimated the fraction of energy loss by charge splitting by performing a Monte Carlo (MC) simulation based on a formalism similar to \cite{dennerl2012a}. For symmetry reasons, we only considered a matrix of $2 \times 2$ pixels and restricted our study to the square region defined by the four pixel centers \citep[see Fig.~4 of][]{dennerl2012a}. Given an energy between 0.1 to 10~keV, we randomly drawn 100 000~photons in the restricted region. At each interaction position, we then assumed the following Gaussian-like function $f(r) = \exp{(-(r/a)^2)}$ to model the radial charge distribution and determine the deposited energy in surrounding pixels. Note that for the MXT pnCCD detector design, the size of the charge cloud distribution evolves by only $\sim$10\% for a photon energy of 1 to 10~keV. For this reason, the pattern ratio and the CS loss depends mainly on the ratio between the photon energy and the fixed LLT value considered for all energies.
        Given the LLT value and the simulated charge distributions in the pixels, we derived the expected pattern statistics and the average energy loss for the four multiplicity at each incident energy.

\section{Experimental results and spectral performance}
\label{sec:experimental_results_and_spectral_performance}
    \subsection{Dark noise and low-level threshold}
    \label{sub:dark_noise_and_low_level_threshold}
    \begin{figure*}[t]
        \centering
        \includegraphics[width=0.85\hsize]{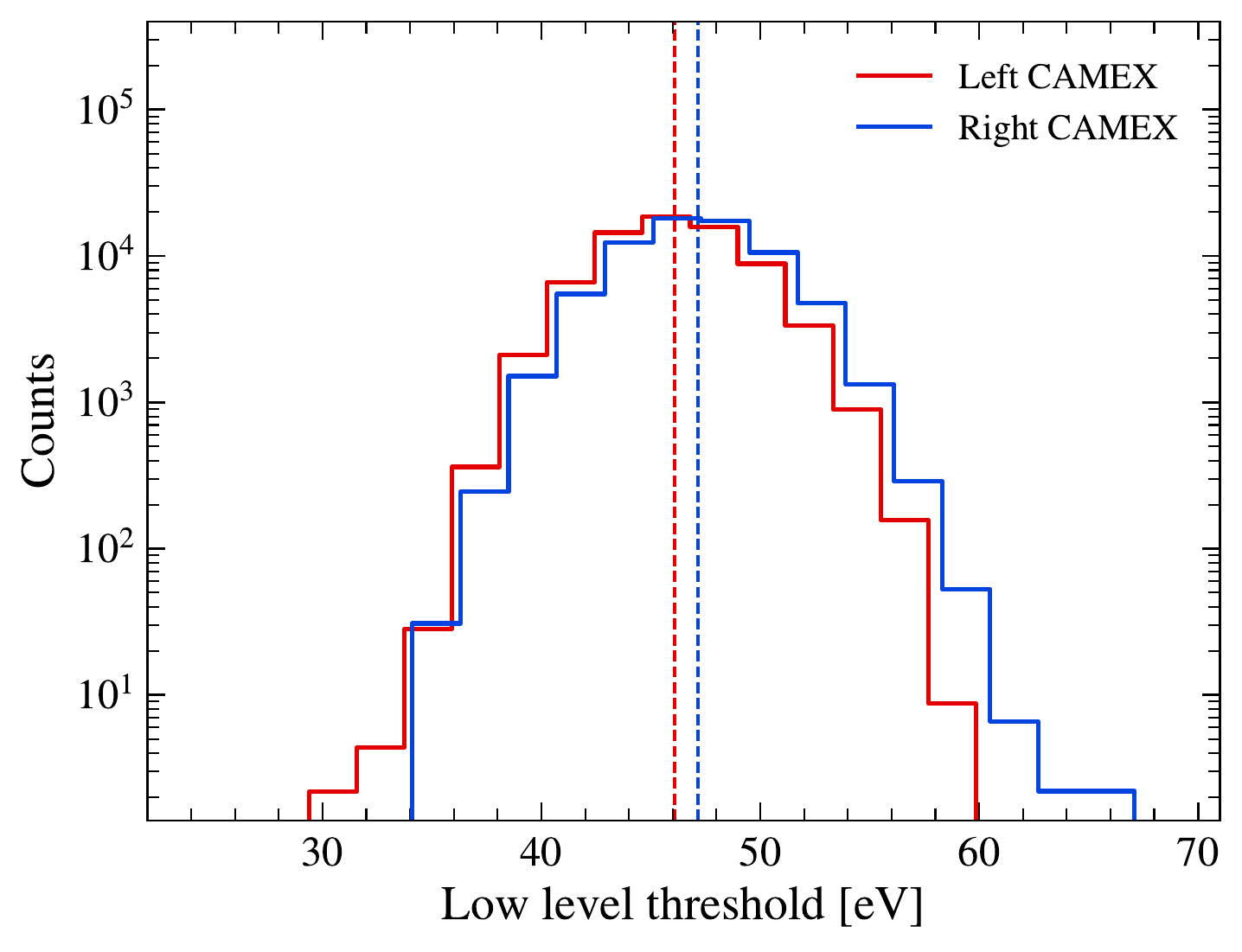}
        \caption{Histogram of LLT values at $-65^{\circ}$C for the $256 \times 256$ pixels of the MXT flight model. Red (blue) color shows the distribution for the left (right) CAMEX. The dashed line indicates the mean value of each histogram. The mean LLT value of the 65536 pixels is $\sim $46.5~eV and corresponds to an average equivalent noise charge of $\sim$$
         3.2 \ \mathrm{e}^{-}_{\mathrm{rms}}$.}
        \label{fig:LLT_map}
    \end{figure*}
    The dark current of a detector, often caused by thermal generation of electrons, represents the baseline value to distinguish valid events from the noise.
    To find X-ray events in a frame, the standard approach is based on the identification of the pixels with ADU values larger than $k$ times their noise level as determined from dark frames. The $256 \times 256$ pixel noise levels are calculated on-board by the MDPU. The algorithm determines the standard deviation of each pixel using typically 200 dark frames, subtracted from the offset and common noise, and where outliers (e.g. cosmic rays) were rejected.
    As discussed in Section~\ref{sub:on_board_data_pre-processing} for the ‘event mode’, this pre-processing is performed in real-time by the FEE. Only pixels with a value above four times the pixel noise level (i.e., LLT value) are considered as valid X-ray events and sent to the ground. This value of $k$ was identified as the best compromise between the event extraction and the spectral performance. \\
    As the offset and LLT might evolve with environmental conditions and aging in-flight, the offset and LLT tables will be regularly calculated (in the MDPU) and updated (in the FEE) on-board. 
    Throughout the Panter campaign, about 40 offset and LLT tables were calculated between $-75^{\circ}$C and $-65^{\circ}$C and showed a very good stability over time. An example of histogram LLT values at $-65^{\circ}$C is shown on Fig.~\ref{fig:LLT_map}. The histogram shows the $256 \times 256$ LLT values assigned to each pixel. It reveals the good uniformity for all pixels with a mean LLT value of 46.61~eV and demonstrates for dark frames the very low noise performance with an equivalent noise charge (ENC) of $3.22\ \mathrm{e}^{-}_{\mathrm{rms}}$ of the MXT detection chain. We can also see that the mean LLT values are slightly different for the two CAMEXs by $\sim$1~eV. We also note that for each CAMEX, the LLT values are not strictly constant over the channels, but decreases by about 1~eV from the left to the right CAMEX column. Finally, we measured the dependence of the LLT table with temperature, and found that the mean LLT value evolves from 45.8~eV at $-75^{\circ}$C to 46.6~eV at $-65^{\circ}$C which is likely due to the higher thermal leakage current.

    \subsection{Initial energy calibration}
    \label{sub:initial_energy_calibration}
    \begin{figure*}[t]
        \begin{subfigure}[b]{0.5\hsize}
            \centering
            \includegraphics[width=\hsize]{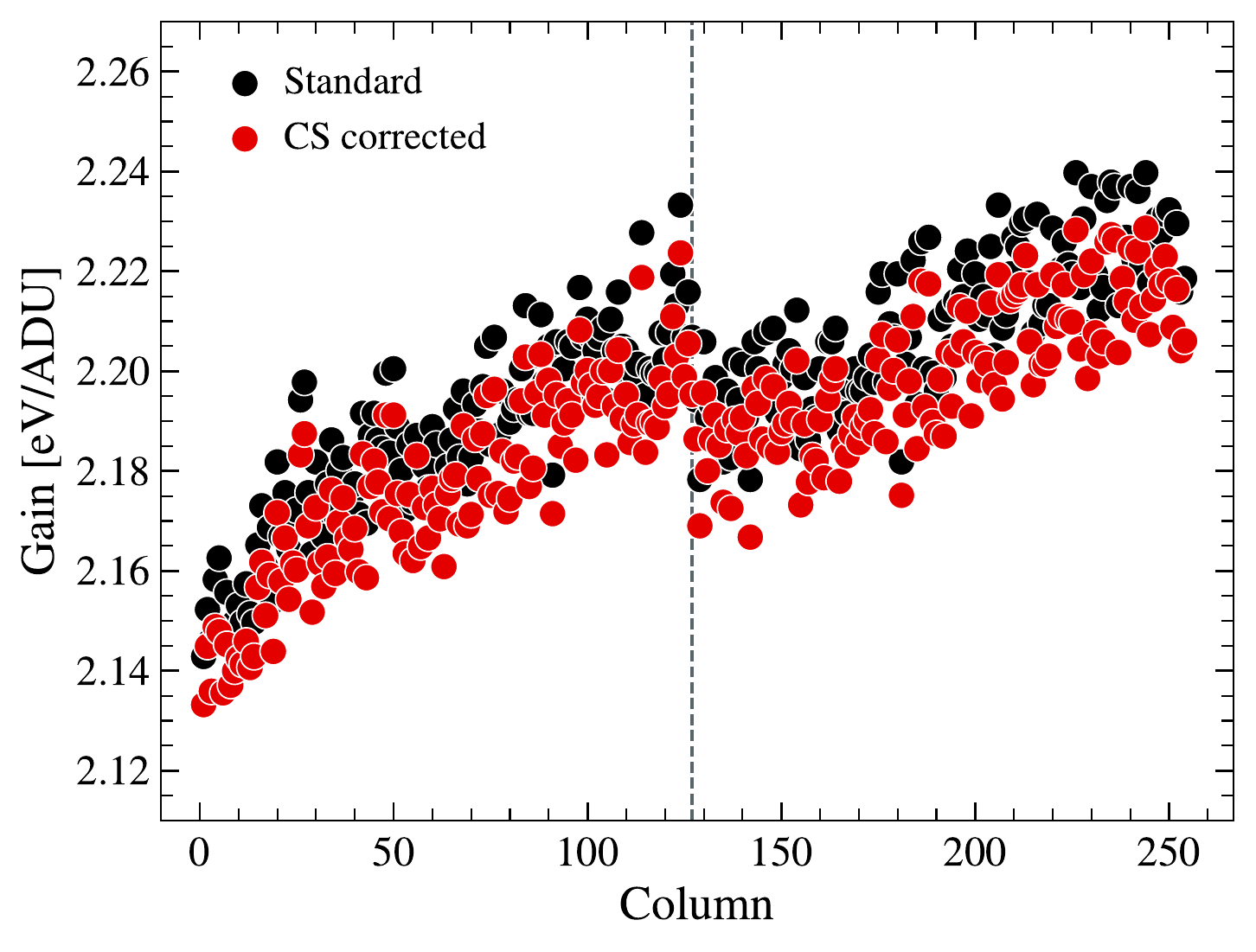}
        \end{subfigure}
        \hfill
        \begin{subfigure}[b]{0.5\hsize}
            \centering
            \includegraphics[width=0.98\hsize]{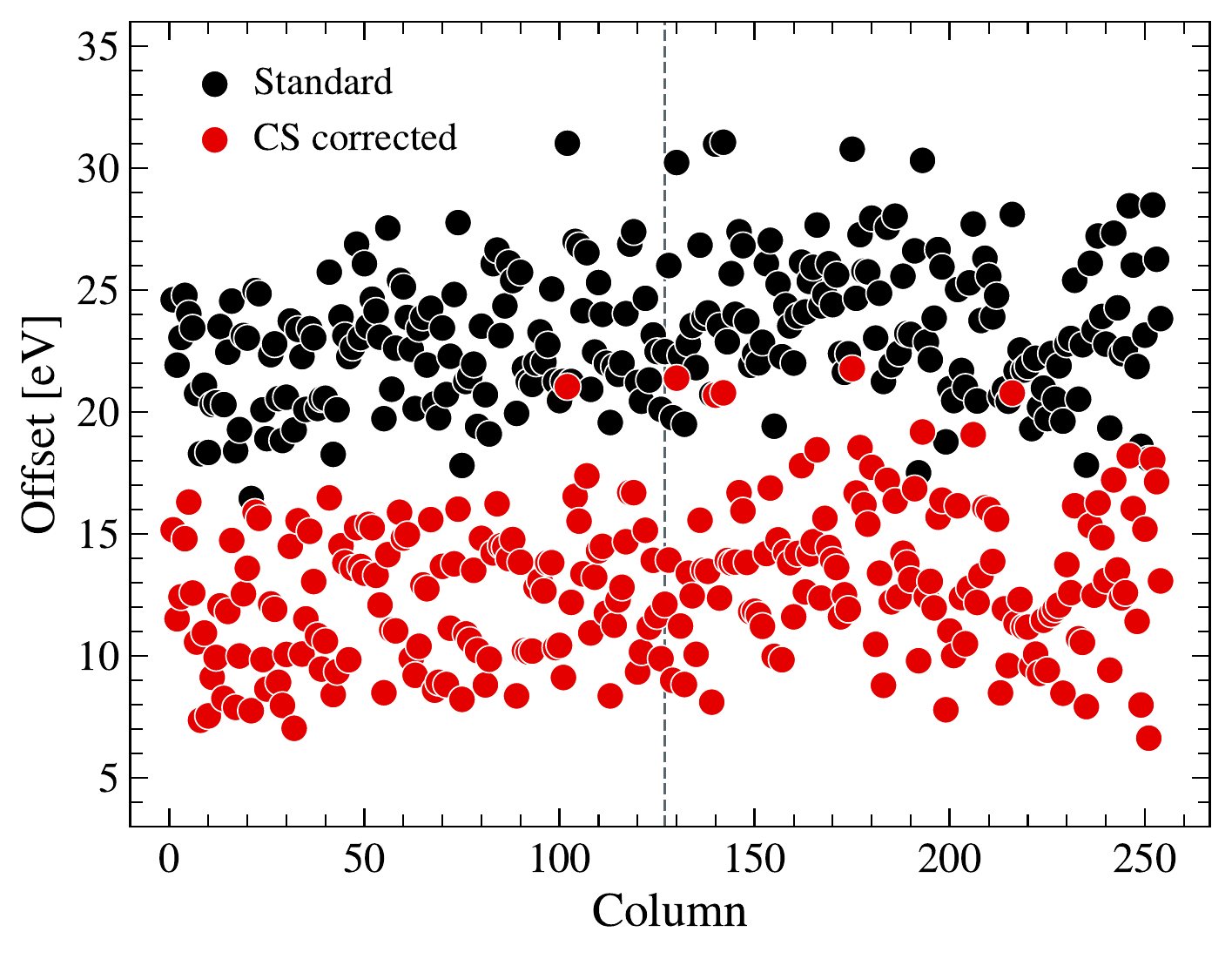}
        \end{subfigure}
        \caption{Detector gains (left panel) and offsets (right panel) as a function of columns (CAMEX channels). Black circles are obtained by the standard calibration method and red circles after considering the charge sharing loss correction in the calibration process. The grey dashed line shows the separation between the left and right CAMEXs.}
        \label{fig:gain_offset}
    \end{figure*}
        The initial energy calibration follows the standard approach described in Section~\ref{ssub:calibration_method} \citep[see also][]{andritschke2008a,ceraudo2020a} where no specific corrections are applied to the raw data (ADU) before running the first calibration process. We extracted only single events and created one spectrum by column with all the lines listed in Table~\ref{tab:energy_lines} and stacked data from positions 9 to 12. We then fed ECC with the 256-column spectra to determine the set of parameters (i.e., gain and offset) per column. The gain and offset distributions are shown in Fig.~\ref{fig:gain_offset}. We found a very good uniformity with a low dispersion of the gain and offset values over the columns. We determined a median gain and offset of about 2.2~eV/ADU and 26~eV, respectively. The dispersion is less than $\sim $1\% for the gains and about 10\% for the offsets. We observed a discontinuity in the gain estimates between the two CAMEXs (separated by the grey dashed line in Fig.~\ref{fig:gain_offset}) which is caused by mismatch variations during the manufacturing processes. We also noticed that the gains gradually increase in the multiplexing sequence direction, from column 0 to 127 for left CAMEX and from 128 to 255 for right CAMEX. We interpreted that as a possible effect of the bandwidth limitations in the FEE analog channels. \\
        To test the accuracy of our energy calibration, we computed the energy scale (i.e., $E_{\mathrm{reconstructed}} - E_{\mathrm{incident}}$) as a function of the energy for single and all events (singles, doubles, triples and quadruples). Figure~\ref{fig:plate_scale} shows that the energy scale is within the $\pm$20~eV instrument requirement up to $\sim $7~keV and $\sim $3~keV for single and all events, respectively. The calibration error for singles is more uniformly spread around zero while the calibration error for all events is systematically positive. We noted in both cases for the Cu-K line ($\sim $8~keV) that the value is shifted regarding the trends at $E < 6$~keV, suggesting a possible small non-linearity of the electronic chain in the high-energy range of MXT. \\
        We investigated how the Cu-K line could affect the calibration law by running a calibration with a data set excluding this line. The results revealed a good overall agreement, with a small tendency to degrade the energy scale at $E > 5$~keV. This confirmed that Cu-K has only a small contribution to the calibration process, which is explained by its limited weight in the correlation compared to the multiple lines existing at $E < 3$~keV. Moreover, ECC performed a correlation with a synthetic spectrum without any background signal while our data sets present a non-negligible background, especially at $E < 3$~keV (Fig.~\ref{fig:combined_spectrum}). As the calibration could also be affected by this problem, we created a data set by selecting only lines with low background signal and performed a new calibration. The results showed that the background has no significant impact on the calibration law and that the measured spectral resolutions are consistent with the one using all energy lines plus a background. We concluded that the set of calibration parameters remains of good quality and consistent even when considering different energy line configurations.

    \subsection{Correction of the multiple events}
    \label{sub:correction_of_the_multiple_events}
    \begin{figure*}[t]
        \begin{subfigure}[c]{0.5\hsize}
        \centering
        \includegraphics[width=\hsize]{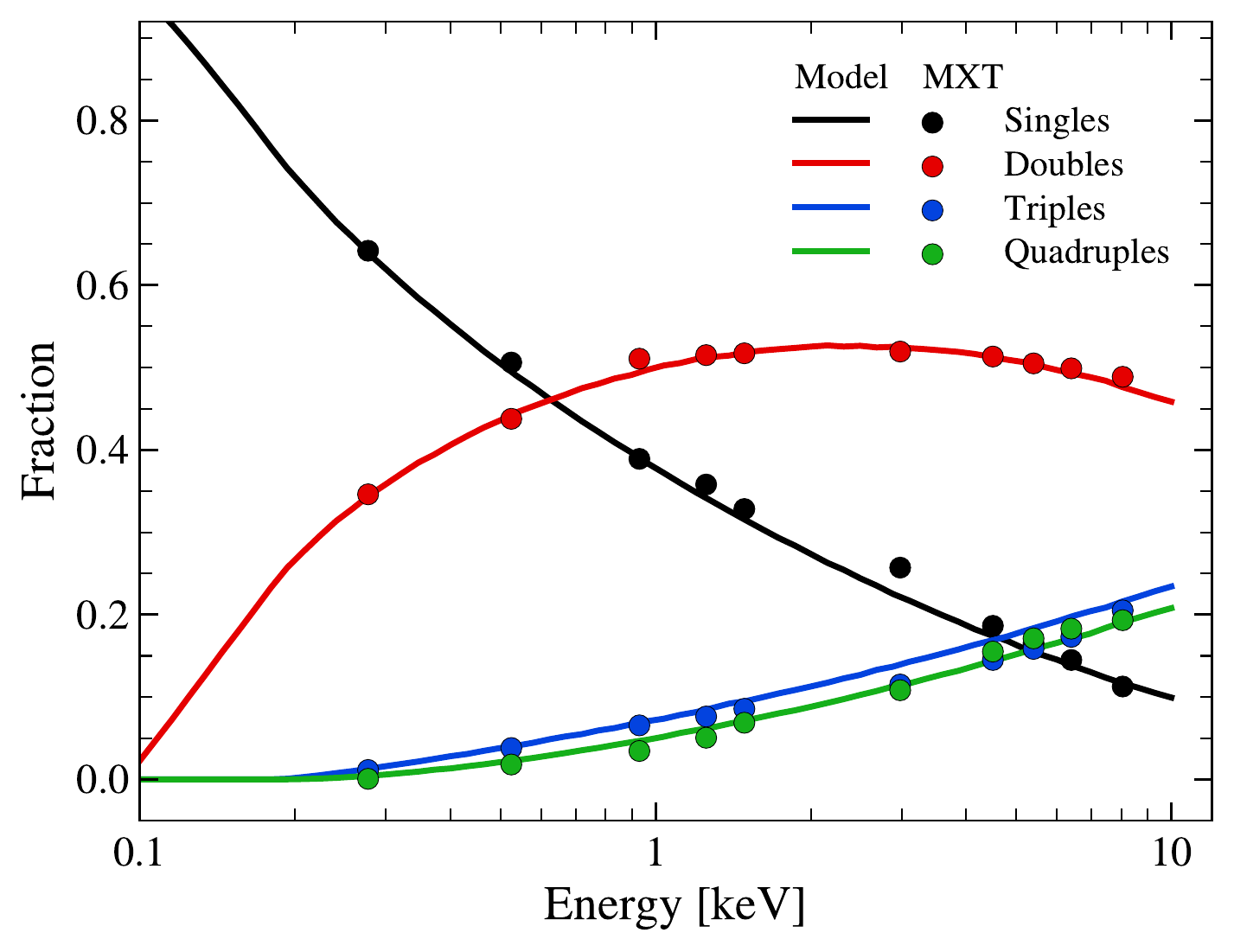}
        \end{subfigure}
        \hfill
        \begin{subfigure}[c]{0.5\hsize}
        \centering
        \includegraphics[width=\hsize]{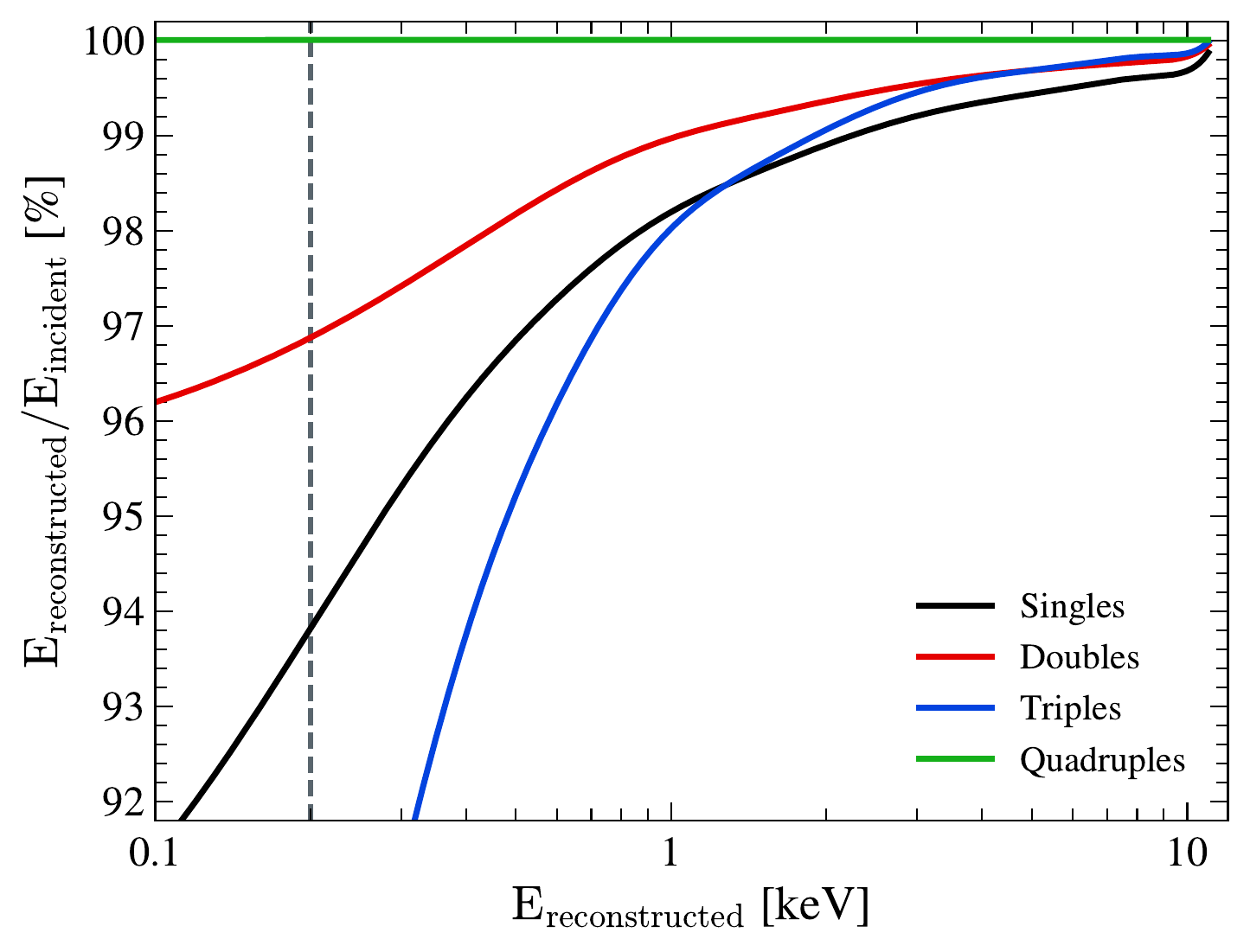}
        \end{subfigure}
        \caption{
        Left panel: Pattern fractions of single (black), double (red), triple (blue) and quadruple (green) events as a function of energy. The circles are ratios determined from the datasets collected at Panter with the MXT flight model. The curves are simulated fractions using \cite{dennerl2012a} formalism and simulated to MXT configuration ($-65^{\circ}$C, 46~eV LLT, 75~$\upmu$m pixel size).
        Right panel: Fraction of charge sharing loss as a function of energy for the four pattern multiplicity produced by an X-ray photon. The dashed grey line represents the 0.2~keV minimum energy threshold of MXT.
        }
        \label{fig:pattern_stats}
    \end{figure*}
        \begin{figure*}[t]
        \begin{subfigure}[b]{0.5\hsize}
            \centering
            \includegraphics[width=\hsize]{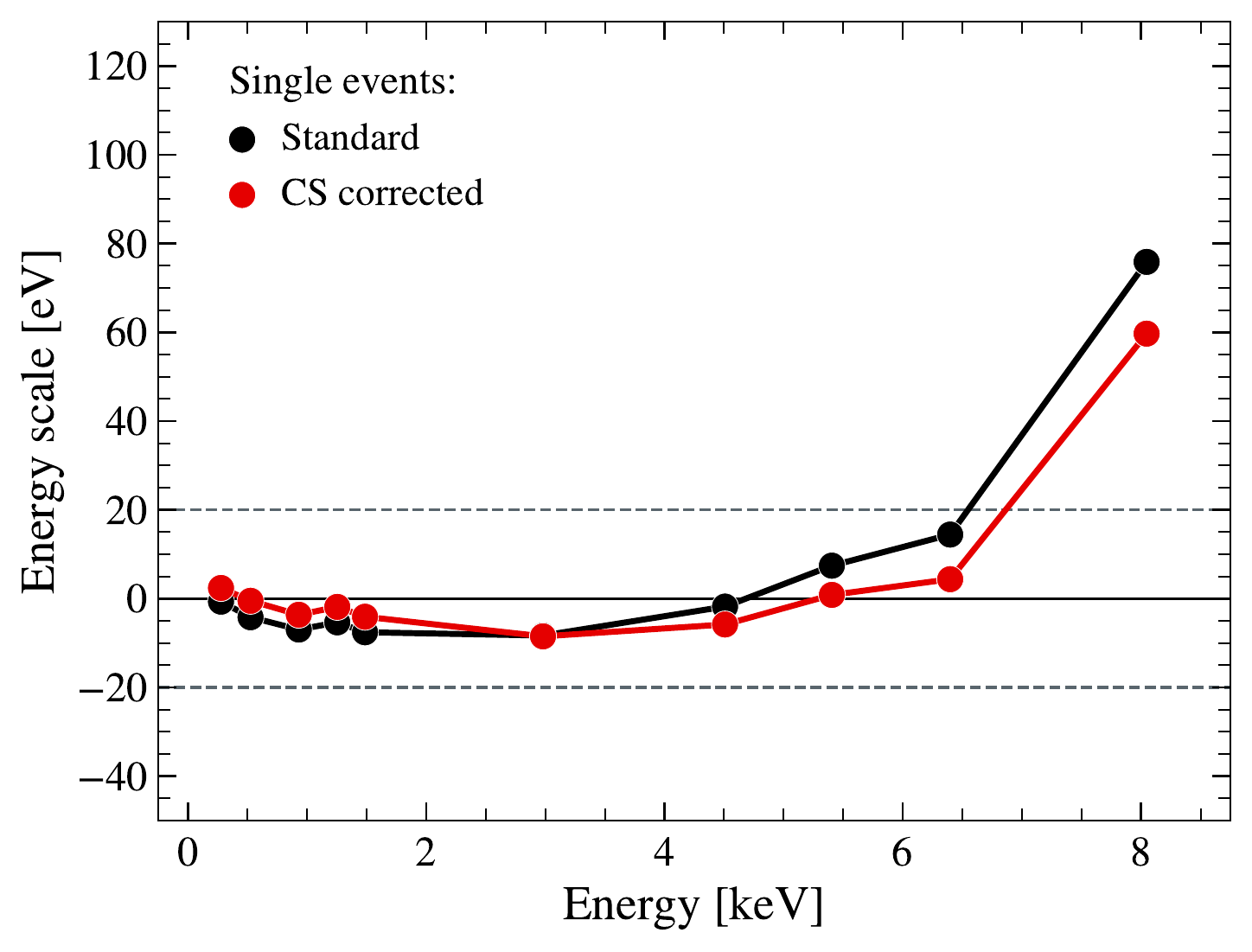}
        \end{subfigure}
        \hfill
        \begin{subfigure}[b]{0.5\hsize}
            \centering
            \includegraphics[width=\hsize]{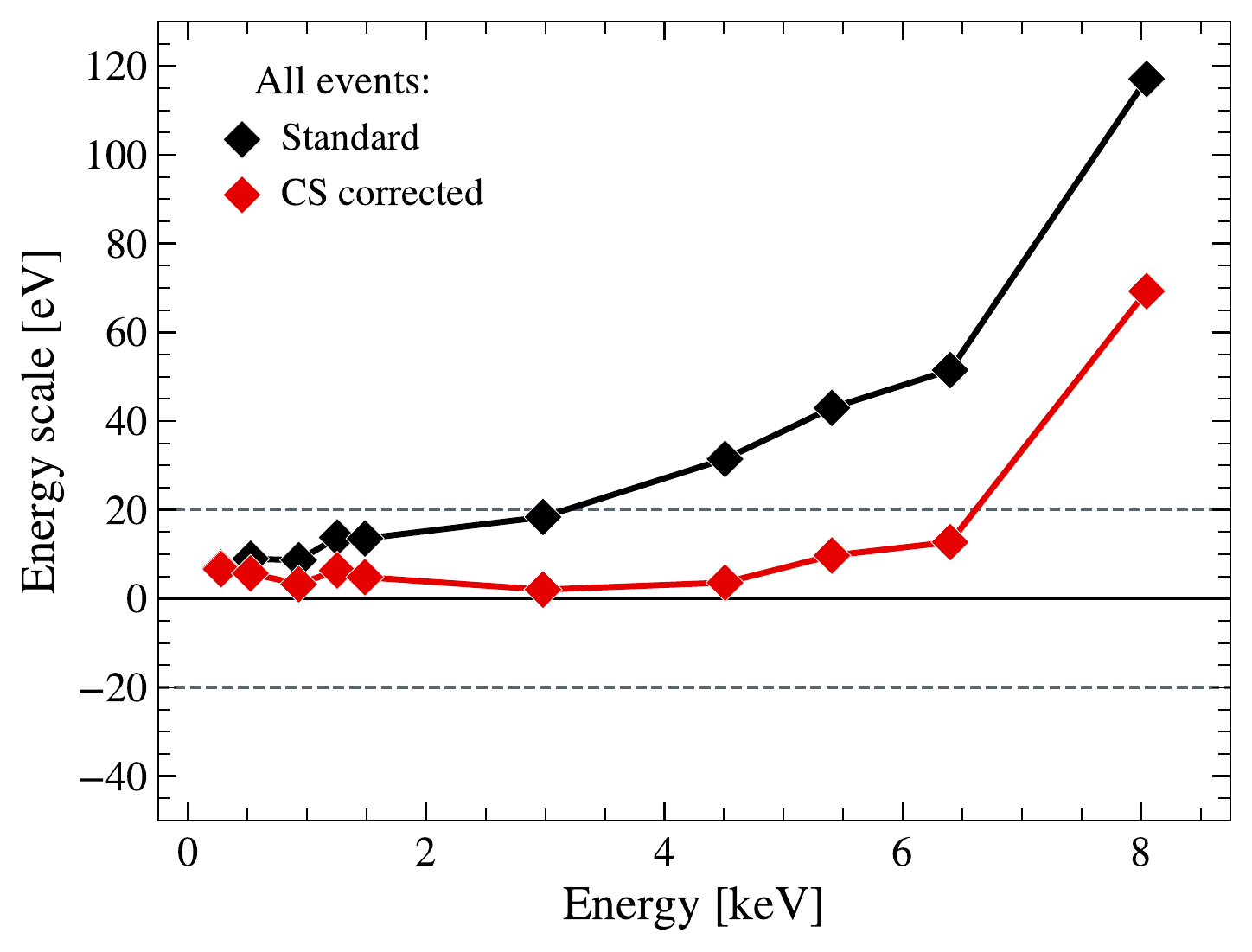}
        \end{subfigure}
        \caption{Energy scale ($E_{\mathrm{reconstructed}} - E_{\mathrm{incident}}$) as a function of energy. Circles (left panel) are for singles and diamonds (right panel) for all events (singles, doubles, triples and quadruples). Black and red colors represents the performance without and with the CS correction. The dashed grey line shows the instrument requirement.}
        \label{fig:plate_scale}
    \end{figure*}
    We ran the CS model described in Section~\ref{sub:management_of_multiple_events} using parameters tailored to the MXT configuration and adapted to the Panter campaign. We used for the split threshold the mean value derived from LLT tables (i.e., 46 eV). We determined the pattern ratios obtained at Panter by considering all event within $\pm$$3\sigma$ of the energy line, except for peaks with a close second line transition (e.g., Cu-K) where we optimized the value by hand. For the $a$ parameter, which defines the radial distribution of the electron packet in the CS model, we found that $a=0.357$ provides good agreement with the observed pattern ratios (left panel of Fig.~\ref{fig:pattern_stats}). We noted that the fraction of pattern is slightly underestimated (overestimated) for triples (quadruples) at $E > 5$~keV. \\
    The results of the CS model for the ratio of the reconstructed photon energy to the incident photon energy is shown in the right panel of Fig.~\ref{fig:pattern_stats}. We found for single, double and triple events that the reconstructed energy is lower than the incoming photon energy. At 0.3~keV, the CS loss corresponds to 14~eV, 8~eV and 26~eV for single, double and triple events, respectively. Then, the ratios increase with $E$ to reach almost one at 10~keV. We found at 8~keV that the CS loss is about 31~eV for singles, 18~eV for doubles and 14~eV for triples.
    Only for quadruple events the ratio remains equal to one over the entire energy range, meaning that no energy is lost for these events. As mentioned earlier, considering the MXT detector geometry, the charge cloud can only be split on a maximum of four pixels. If a quadruple event is detected, it means that the four pixel values are above the LLT value and that no energy from the incident photon was lost during the FEE thresholding step. \\
    To integrate the correction of the CS effect into the calibration process, the order of the steps applied to the raw data has to be considered carefully. As with CTI, CS loss can be mitigated during the calibration process. Because single raw events are affected by the CS loss, we started by creating a synthetic spectrum for ECC that suffers from a CS loss. It was done by using the relation for singles found in the right panel of Fig.~\ref{fig:pattern_stats}. This allows us to disentangle the possible effect of the CS on the calibration parameters and anticipate the CS correction applied later. The gains and offsets obtained using this approach are shown in red color in Fig.~\ref{fig:gain_offset}. We found that gains and offsets have similar trend to the one without CS correction (Section~\ref{sub:initial_energy_calibration}) but with lower overall values. This is a direct consequence of the spectral shift towards the lower energies of the synthetic spectrum caused by the preliminary CS correction. 
    We then applied the set of parameters derived from single events on raw ADU data (without any CS correction) to obtain calibrated data in keV. Once calibrated, we corrected the reconstructed events from the CS loss by applying the relation derived for each multiplicity visible in the right panel of Fig.~\ref{fig:pattern_stats}. Again, we tested the accuracy of the calibration by measuring the energy scale on single and all events spectra (Fig.~\ref{fig:plate_scale}). For singles, we found that the CS correction slightly improves the position of the lines at $E < 5$~keV and more significantly at $E > 5$~keV. For all events, this additional correction significantly improves the line positions over the entire energy range. The positions fall within the requirement up to $\sim$6.5~keV compared to $\sim$3~keV with the standard approach. In addition, the CS correction contributed to reduce the supposed non-linearity previously observed in the high energy domain of MXT. However, despite the CS correction, we noted that the position of Cu-K is still off the instrument requirement, suggesting that a non-linear calibration would be necessary to improve the positions at the end of the MXT energy band.
    The effect of the CS correction on the energy resolution is further discussed in Section~\ref{sub:energy_resolution}.
        
    \subsection{Charge transfer (in-)efficiency}
    \label{sub:charge_transfer_efficiency}
    \begin{figure*}[t]
        \begin{subfigure}[b]{0.5\hsize}
            \centering
            \includegraphics[width=\hsize]{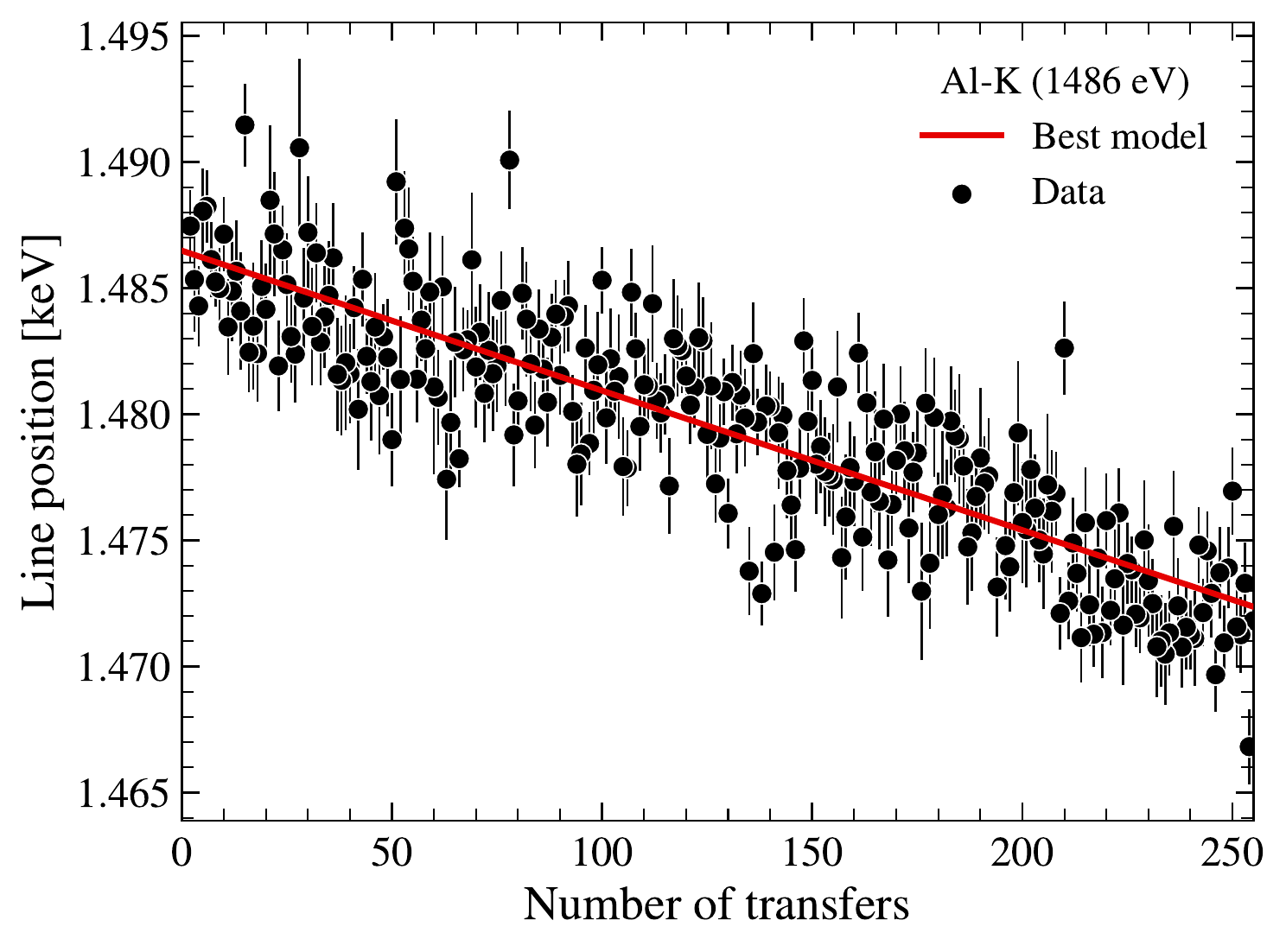}
        \end{subfigure}
        \hfill
        \begin{subfigure}[b]{0.5\hsize}
            \centering
            \includegraphics[width=0.96\hsize]{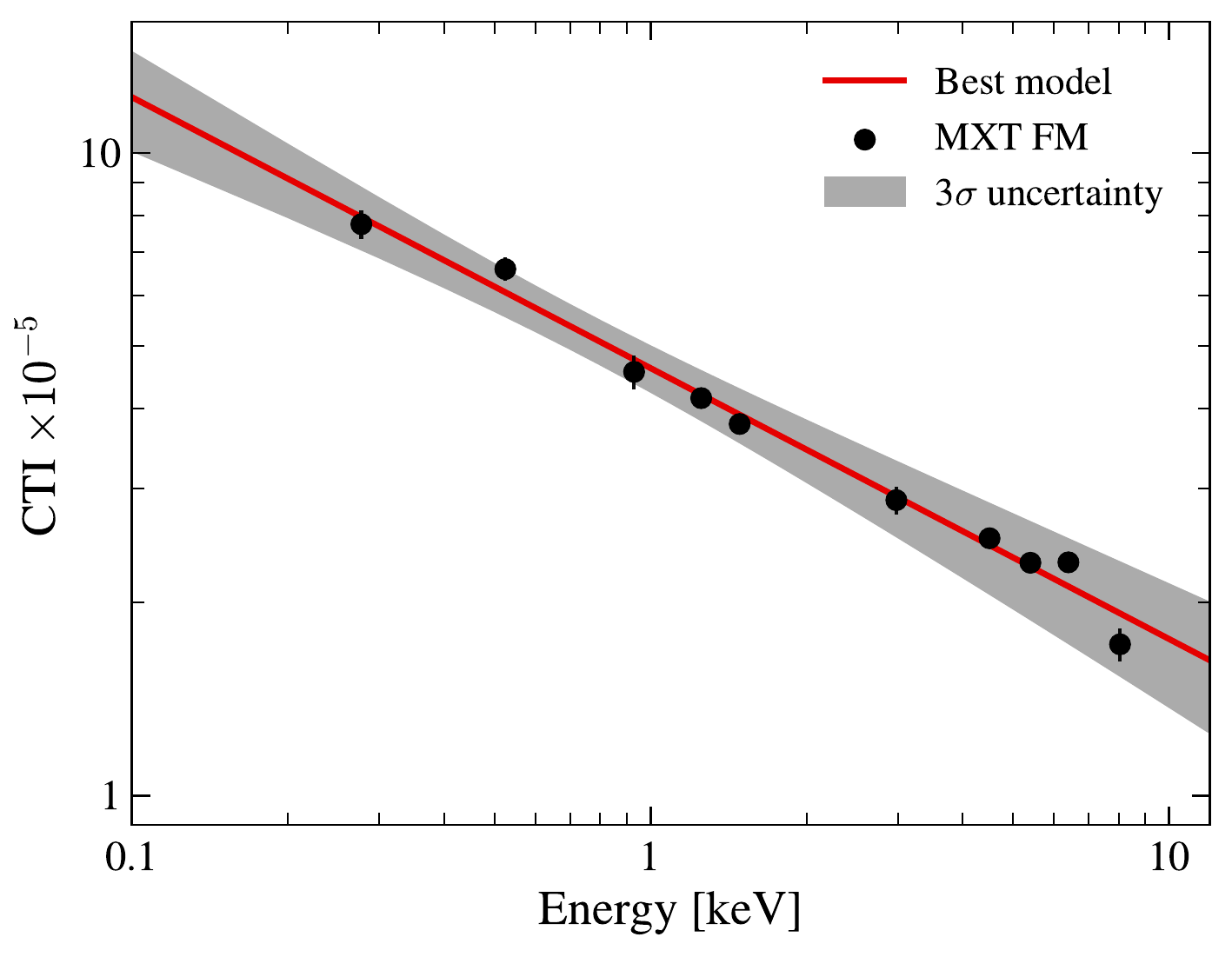}
        \end{subfigure}
        \caption{Left panel: Line center position for Al-K as a function of transfers number. Circles represent peak positions determined by a Gaussian fit on row spectra. The best-fit model obtained to derive the CTI is shown as a red line. 
        Right panel: Charge transfer inefficiency (CTI) as a function of energy. The best-fit model is shown in red. The grey area represents the 3$\sigma$ uncertainty associated to the model.}
        \label{fig:cti}
    \end{figure*}

    The left panel of Fig.~\ref{fig:cti} clearly shows the effect of the CTI on the measured line position of Al-K. For closest rows to the CAMEX (i.e., $\lesssim$5~transfers), only a slightly fraction of the incident photon energy is captured by crystal defects (probably in the frame-store area) leading to a measured position close to the theoretical value of 1.486~keV. Then, as the number of transfers increases, the position is gradually shifted to a lower energy and reaches its minimum at the farthest row from the anode. The visible dispersion of the positions might be explained by the statistical fluctuations of the process that traps and re-emits electrons. To determine the CTI of the MXT detector at the beginning of its life and evaluate its trend as a function of energy, we consider the ten most intense lines obtained at Panter that probe the entire MXT energy range. We then derived the CTI on each line individually as shown in the left panel of Fig.~\ref{fig:cti} and previously described in Section~\ref{ssub:charge_transfer_efficiency_correction}. The right panel of Fig.~\ref{fig:cti} shows the resulting estimates as function of the energy. The CTI is in the range of $10^{-5}$--$10^{-4}$ and we found that it decreases as the incoming energy photons increases. We expect this trend because the number of charges generated by an X-ray photon is proportional to its incoming energy and the electrons captured by crystal defects is a constant number related to the number of traps. It is therefore consistent to have low energy photons with a higher CTI value.
    We modeled the CTI with $E$ using a power law function and the best-fit model is found for: 
    \begin{equation}
    \label{eq:CTI_model}
        \mathrm{CTI}(\mathrm{E}_i) = 4.63 \pm 0.09 \cdot 10^{-5} \times \mathrm{E}_i^{-0.42 \pm 0.02}, \, \mathrm{with\ E_i\ in\ keV.}
    \end{equation}
    We noted that only an iterative process can determine the absolute CTI due to the interdependence of the gain (column-wise) and CTI (row-wise) corrections. (see Section~\ref{ssub:charge_transfer_efficiency_correction}). We observed that after one iteration loop, the CTI estimates are even smaller ($10^{-7}$--$10^{-6}$) and constant over the entire energy range. This confirmed that our first estimates were already close to the absolute CTI. We also investigated the possible effect of the CS loss correction on the CTI estimates. We determined the CTI for both data sets, with and without CS correction. We found very good agreement for the two resulting CTI trends. However, it is worth noting that the CTI should be derived before any CS correction on the data, given that the correction would tend to reduce the dispersion between the line positions of the low and high row spectra. Considering the current CTI, the CS correction has only a negligible effect on the line positions deviation observed in left panel of Fig.~\ref{fig:cti}, resulting in a similar CTI trend between the two data sets. \\
    These CTI estimates represent the current level of the detector defects and impurities before the first radiation damage that MXT will suffer during in-orbit operations. The evolution of the CTI and how space radiation will affect MXT performance have been theoretically investigated using \texttt{Geant4} simulations \citep{ceraudo2019a} and will be experimentally investigated on the flight spare model produced from the same CCD wafer. 

    \subsection{Energy resolution}
    \label{sub:energy_resolution}
    \begin{figure}[hp]
        \begin{subfigure}[b]{0.49\hsize}
            \centering
            \includegraphics[width=0.80\hsize]{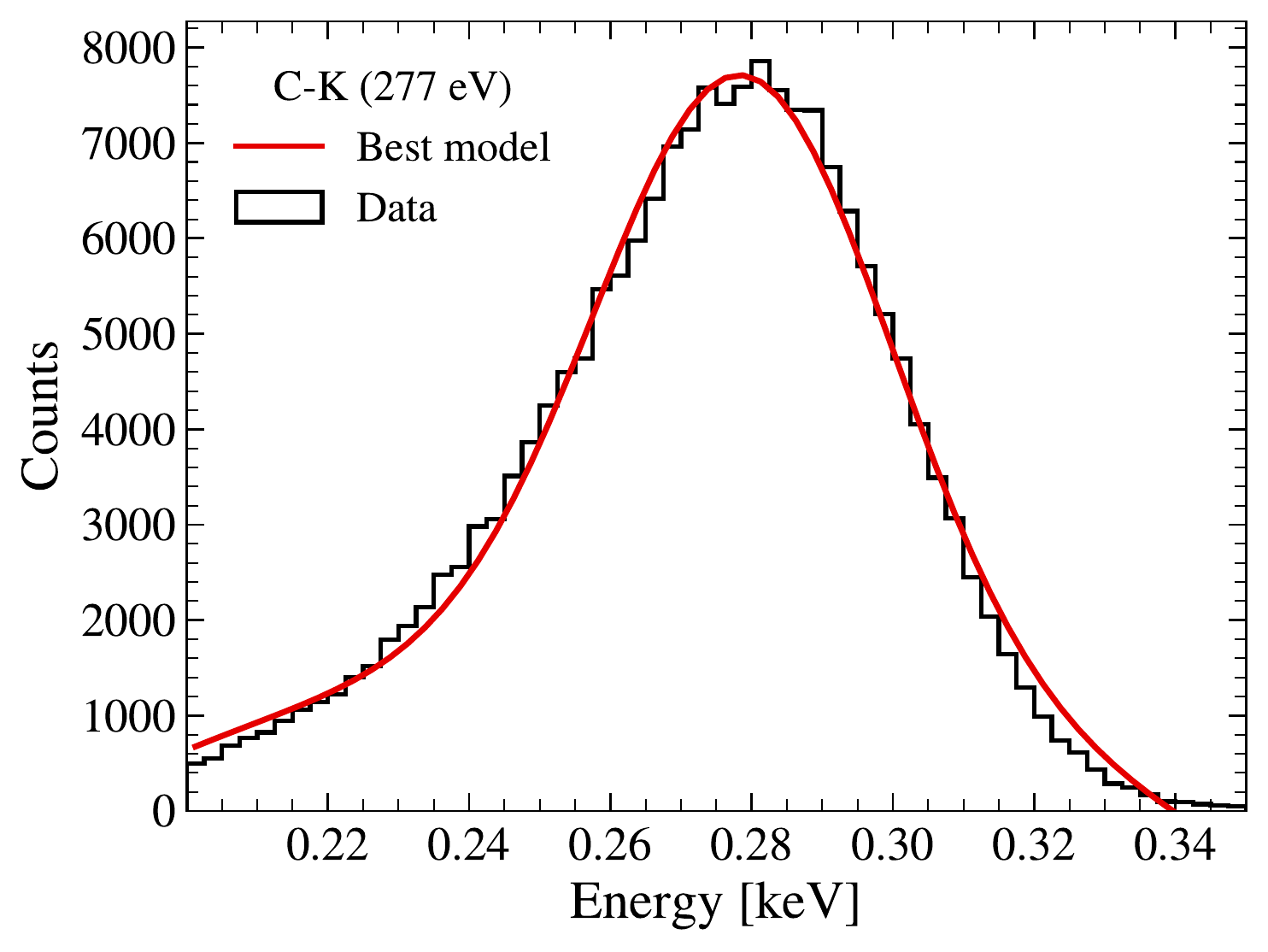}
            \includegraphics[width=0.80\hsize]{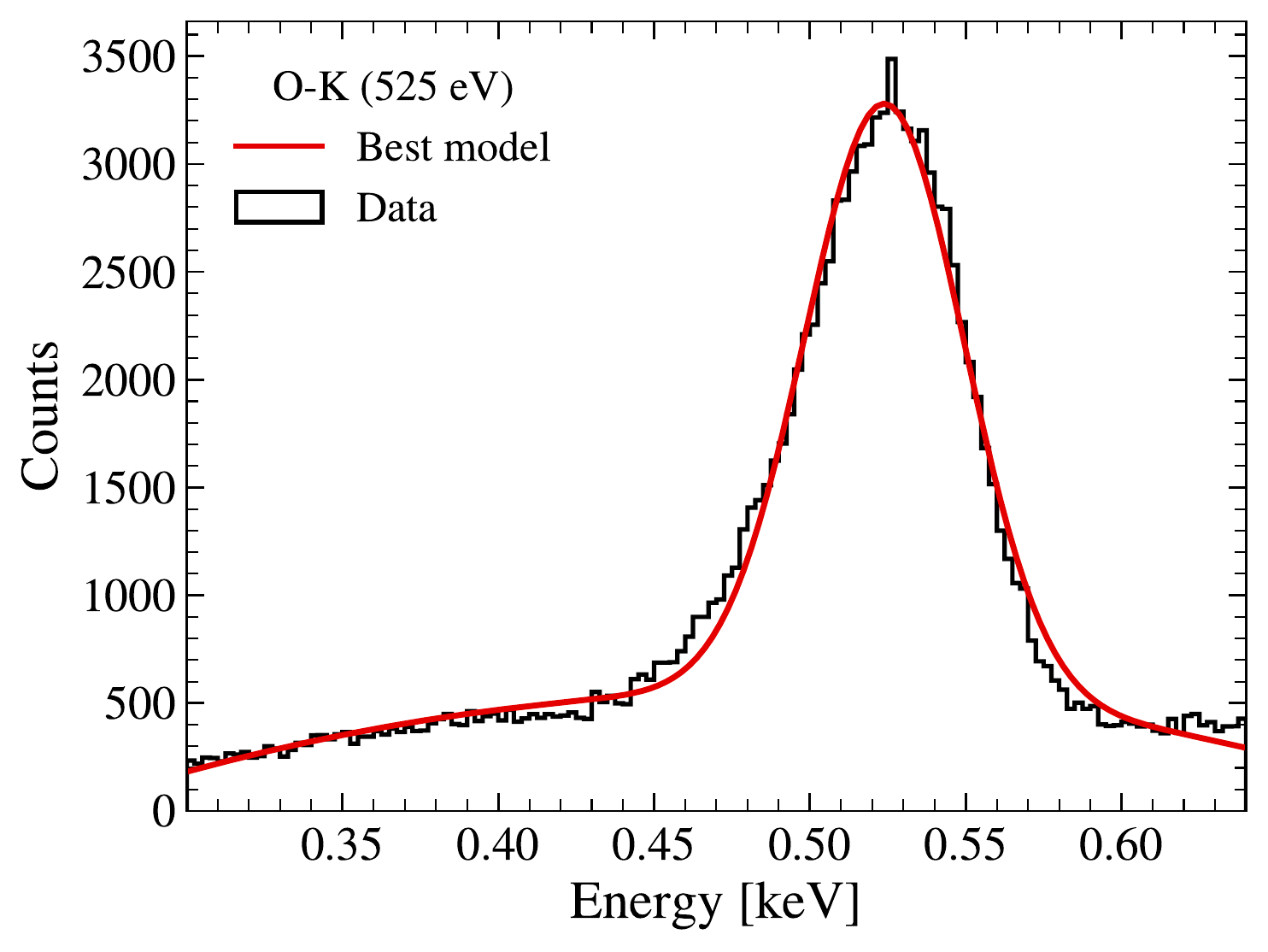}
            \includegraphics[width=0.80\hsize]{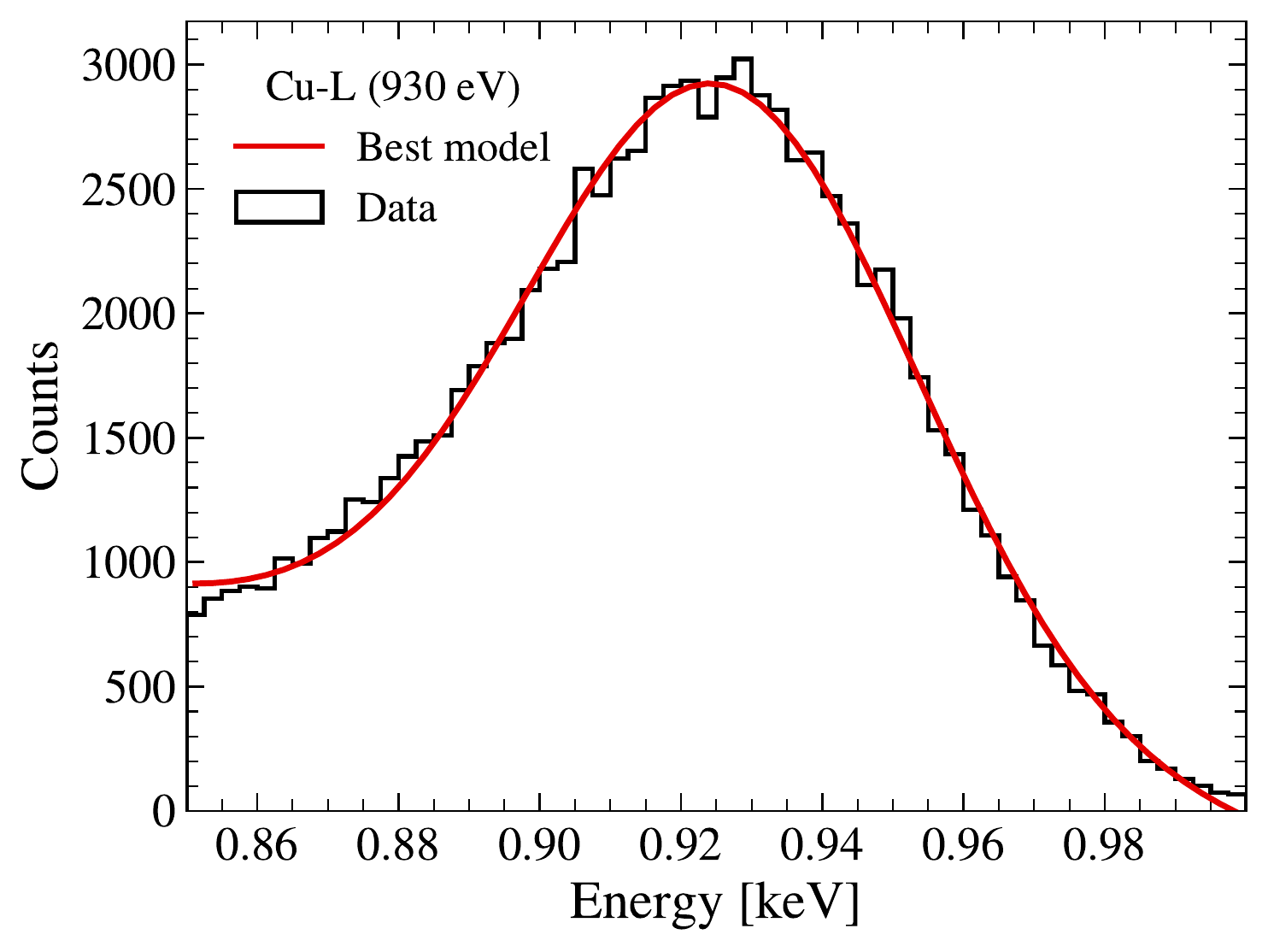}
            \includegraphics[width=0.80\hsize]{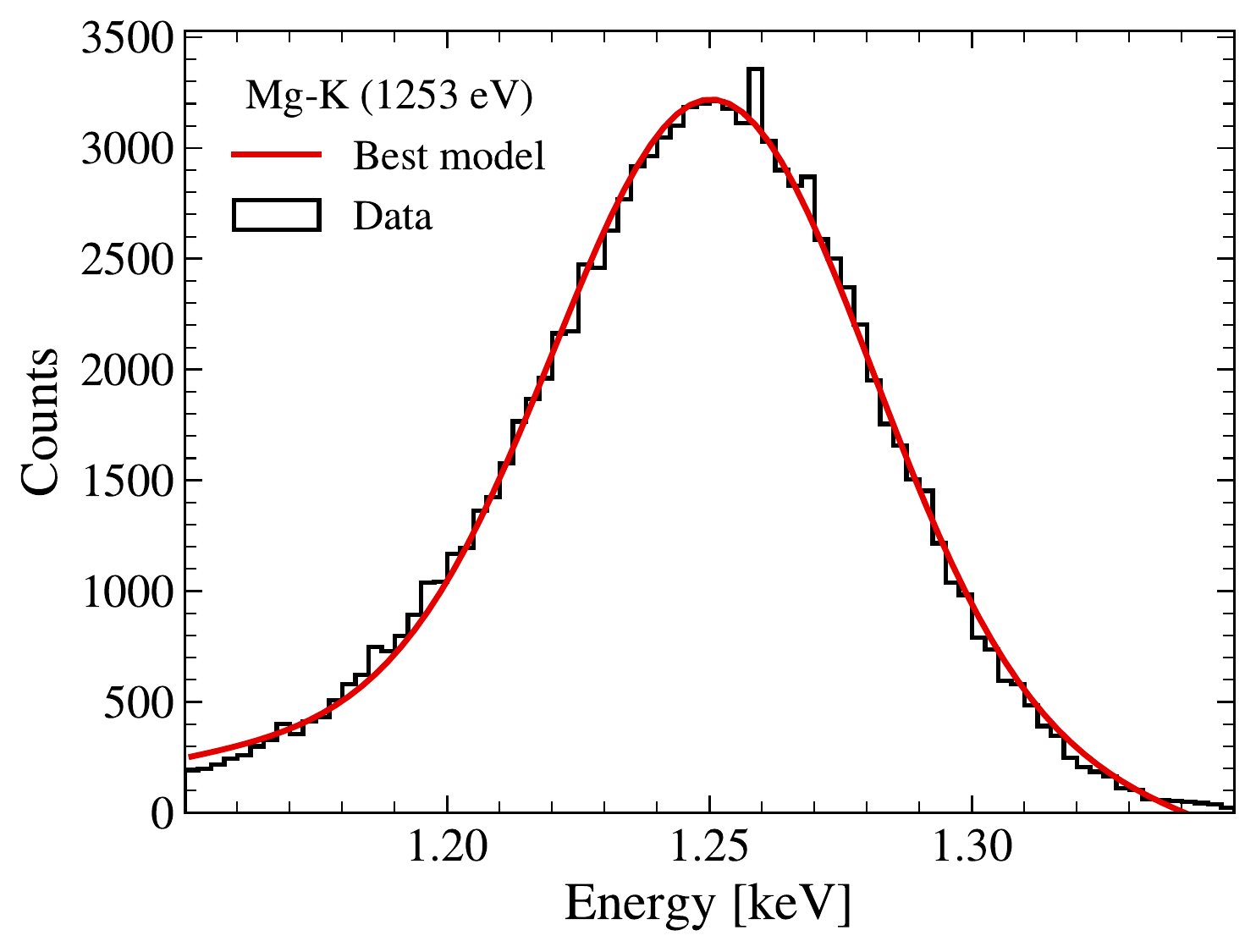}
            \includegraphics[width=0.80\hsize]{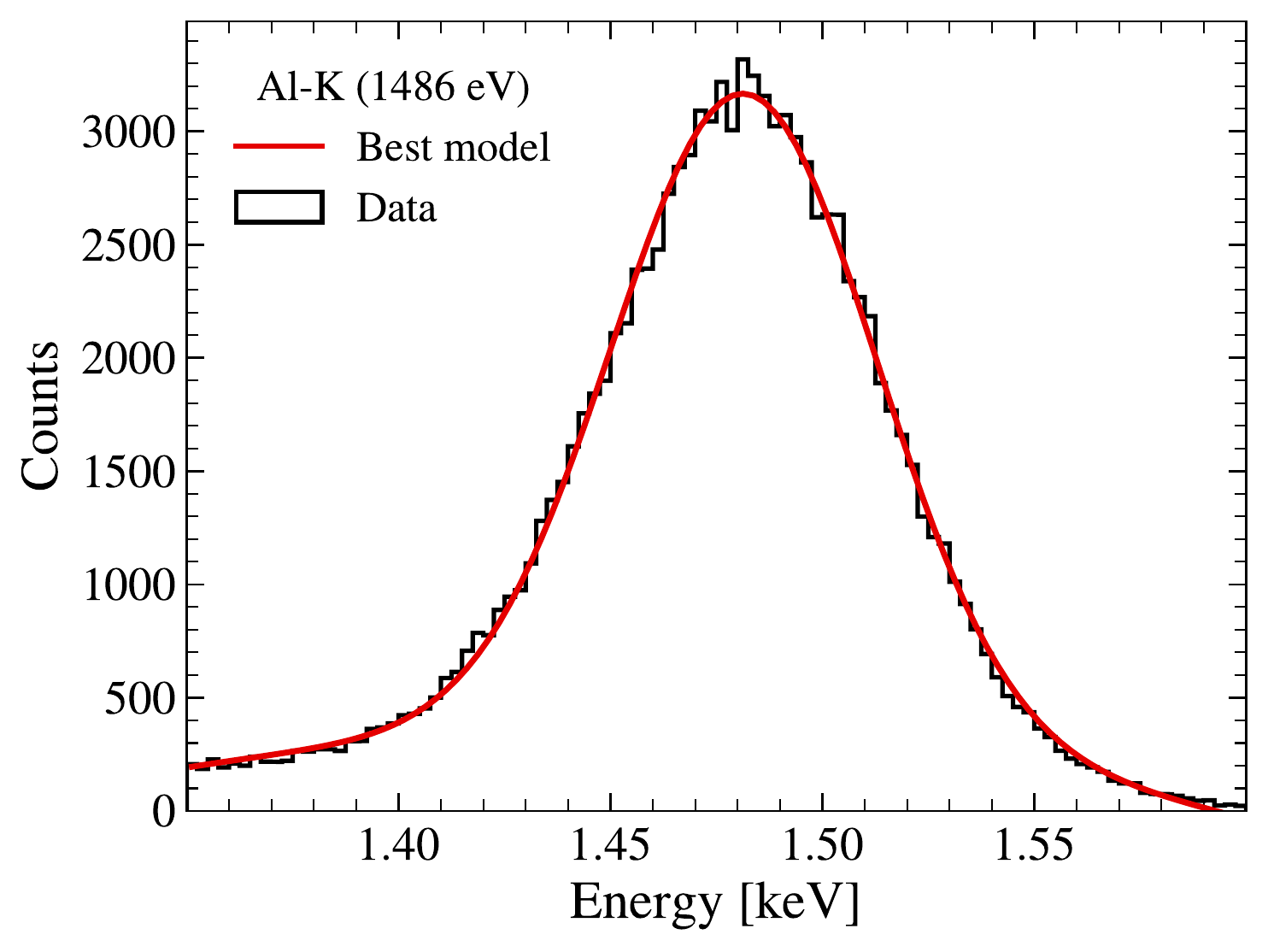}
        \end{subfigure}
        \hfill
        \begin{subfigure}[b]{0.49\hsize}
            \centering
            \includegraphics[width=0.80\hsize]{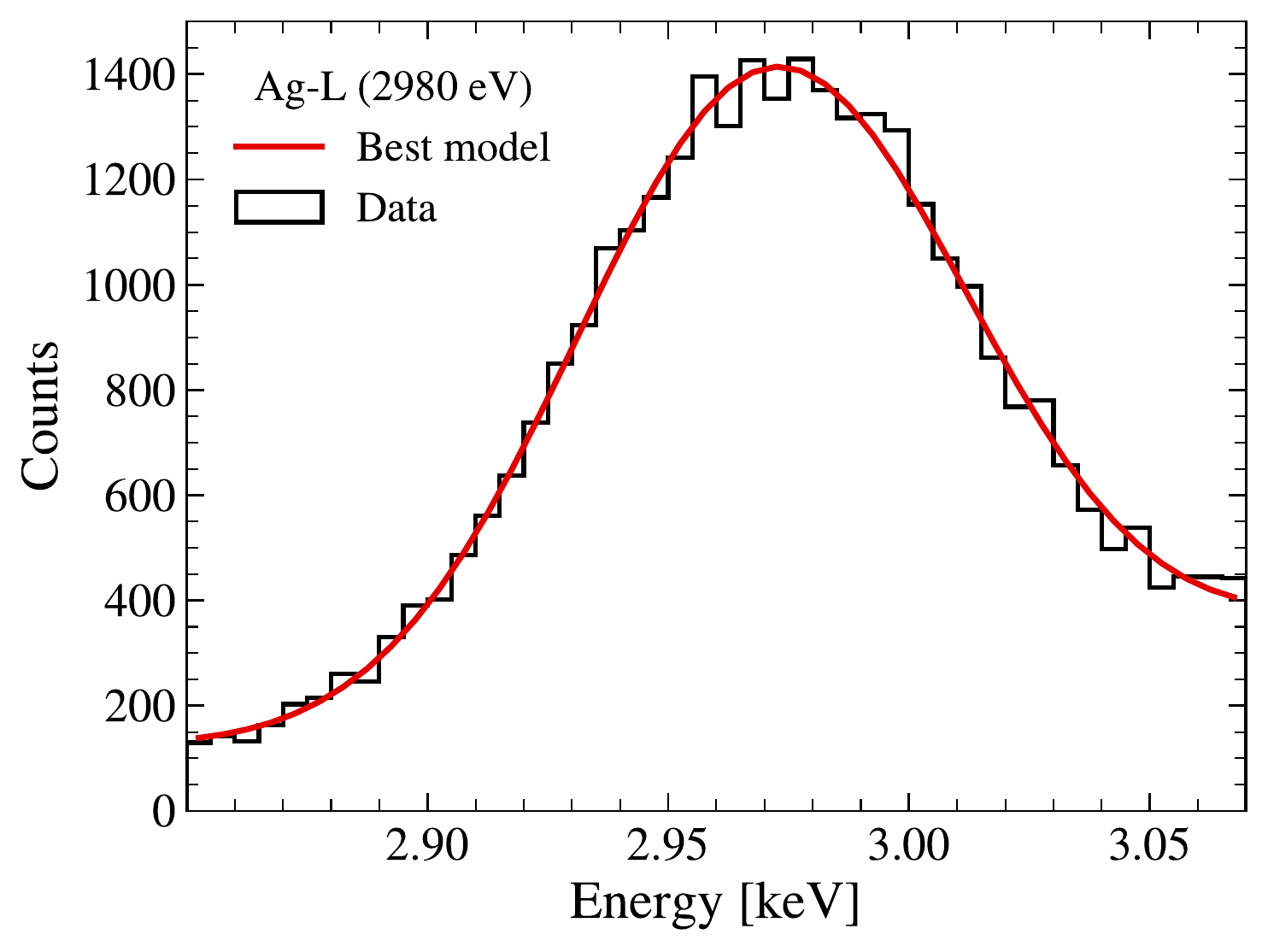}
            \includegraphics[width=0.80\hsize]{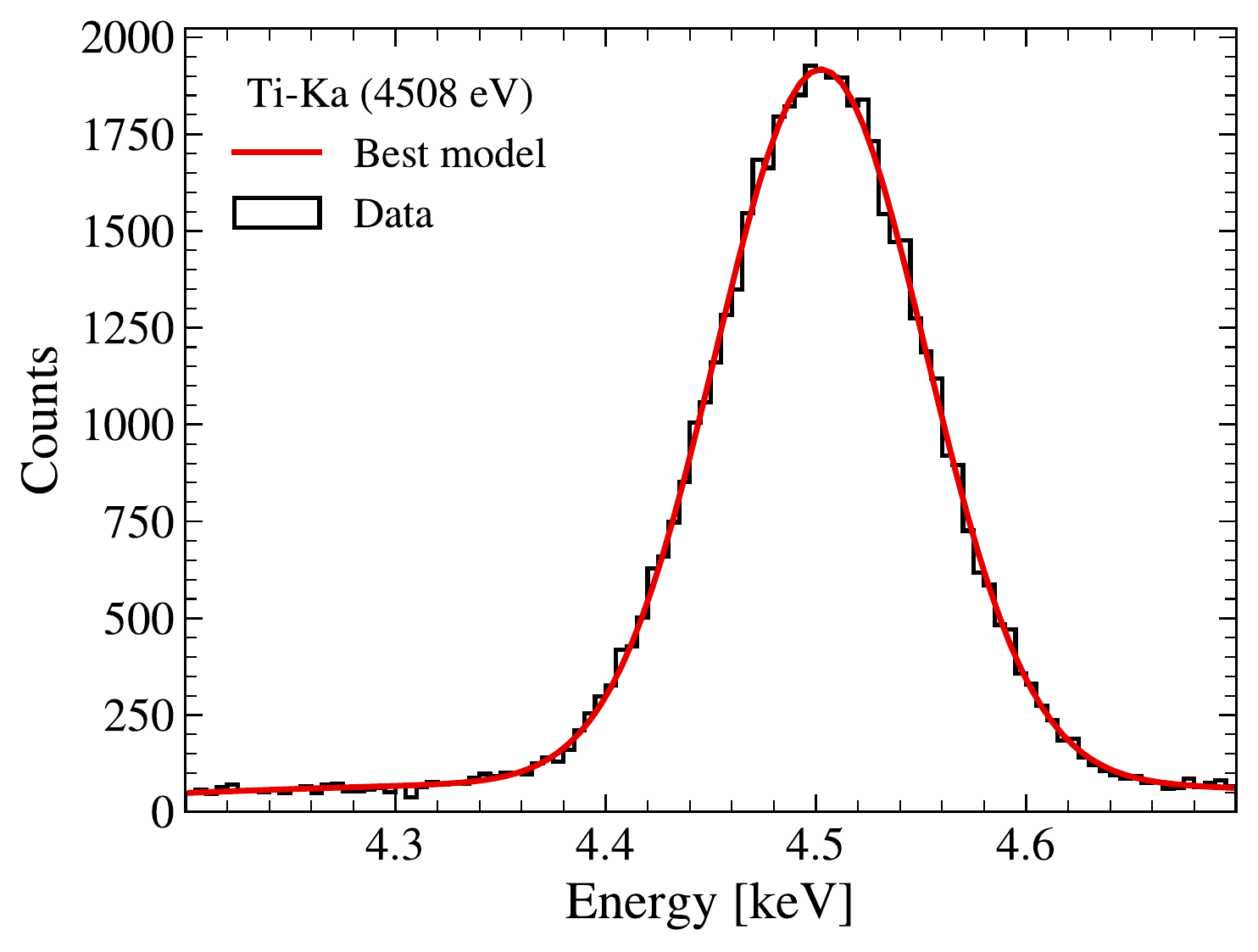}
            \includegraphics[width=0.80\hsize]{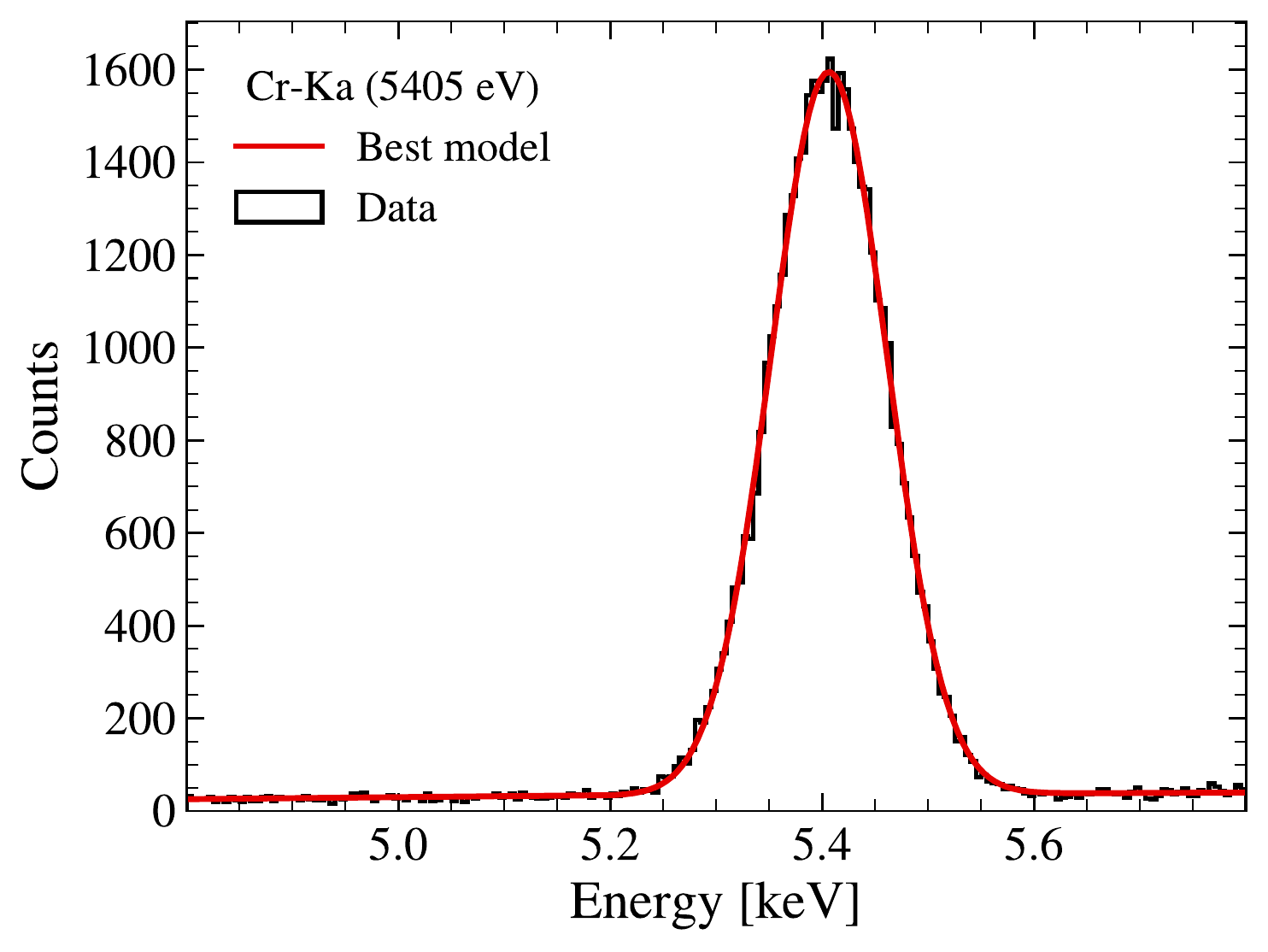}
            \includegraphics[width=0.80\hsize]{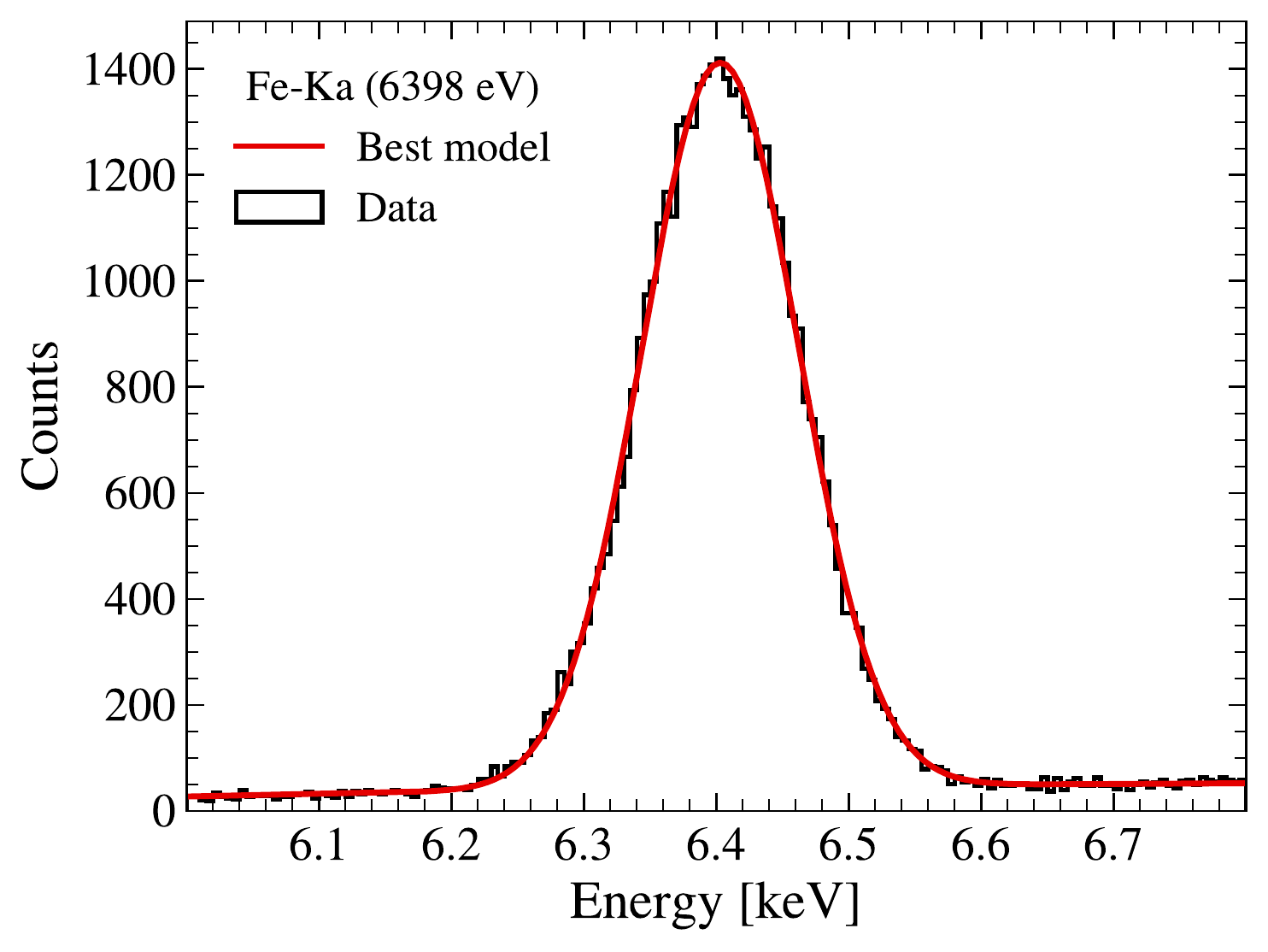}
            \includegraphics[width=0.80\hsize]{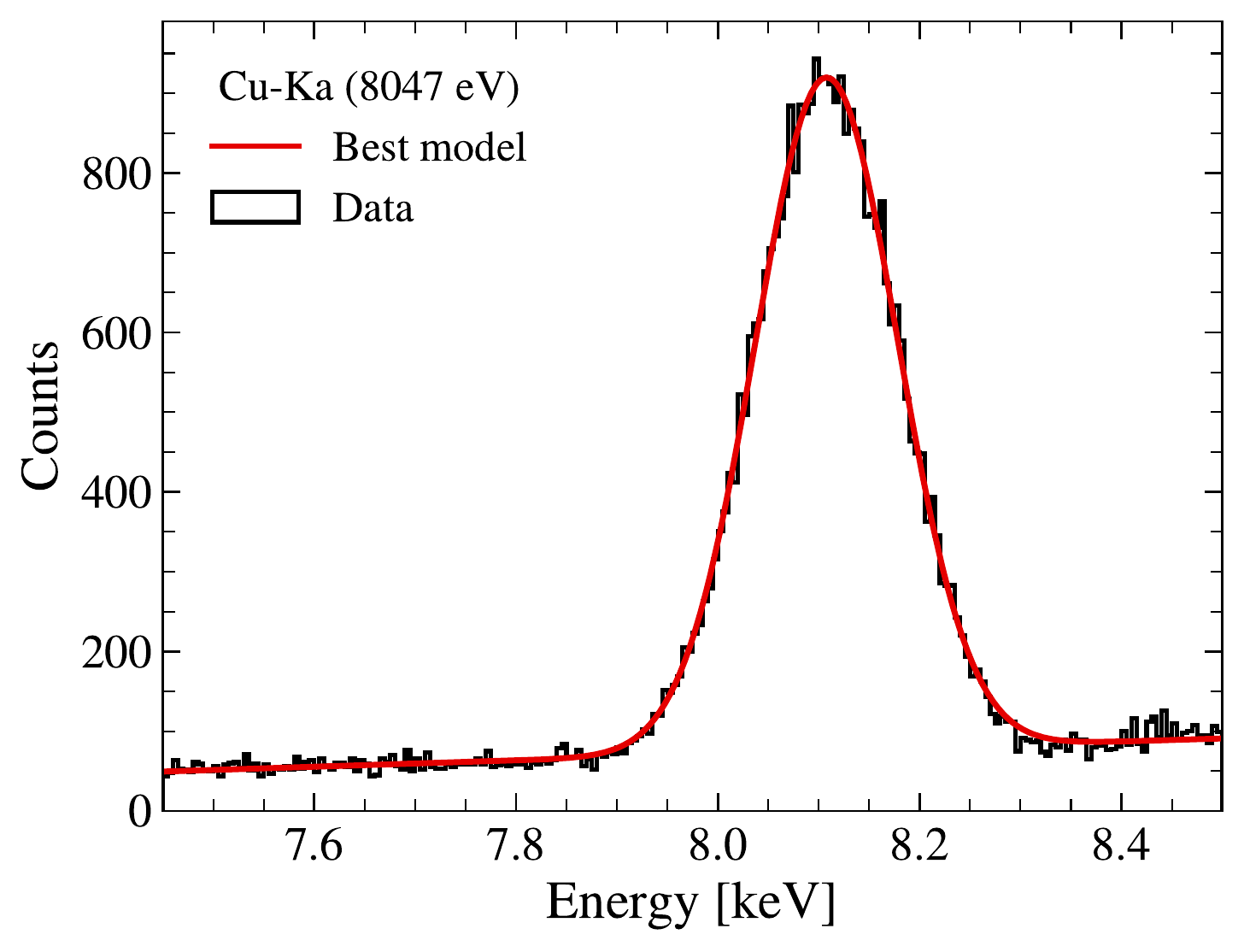}
        \end{subfigure}
        \caption{Spectrum of single events calibrated in energy corrected from charge sharing effect. The red curve is the best-fit model found and used to determine the MXT energy resolution.}
        \label{fig:all_spec}
    \end{figure}

    \begin{figure}[!t]
        \begin{subfigure}[b]{0.5\hsize}
            \centering
            \includegraphics[width=\hsize]{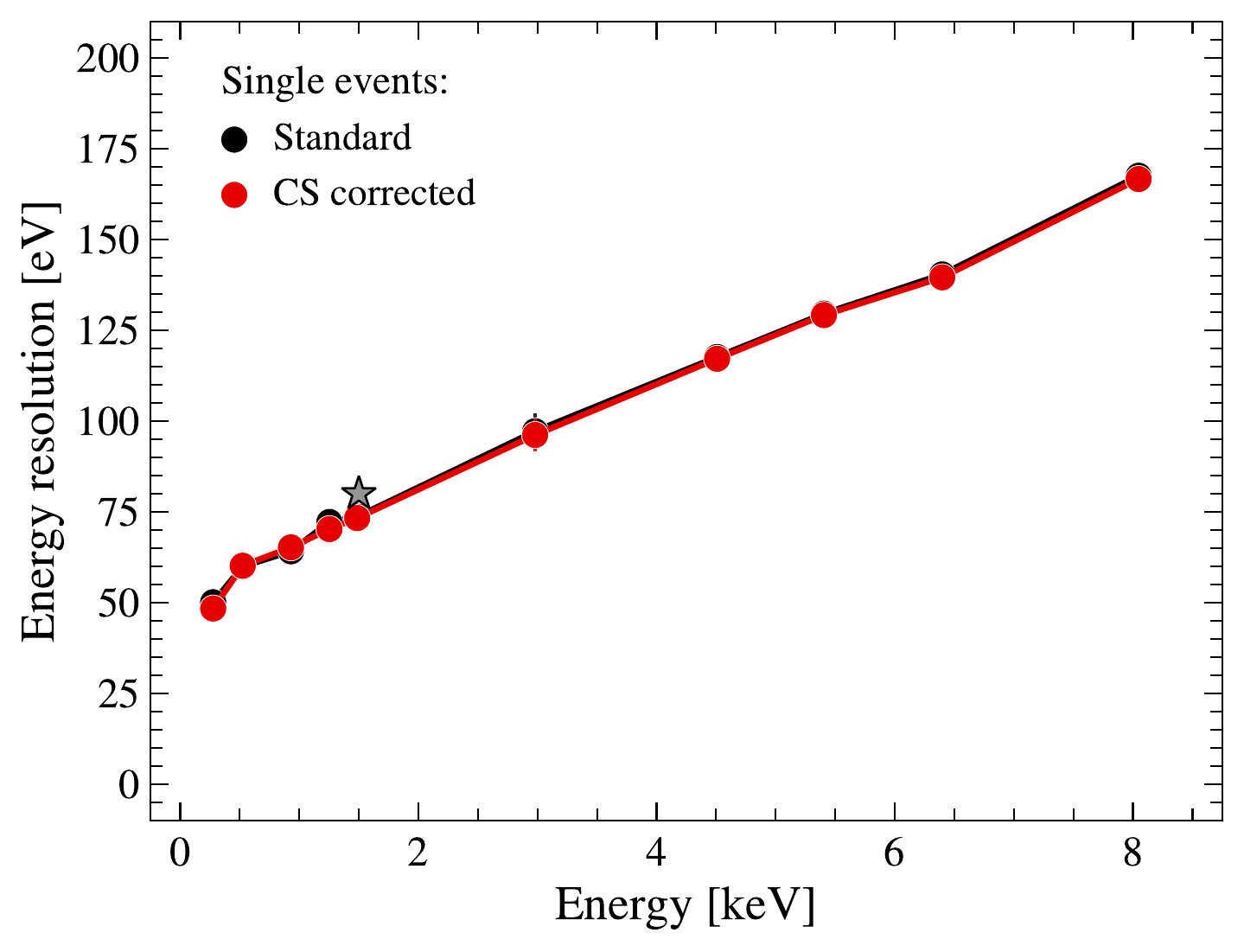}
        \end{subfigure}
        \hfill
        \begin{subfigure}[b]{0.5\hsize}
            \centering
            \includegraphics[width=\hsize]{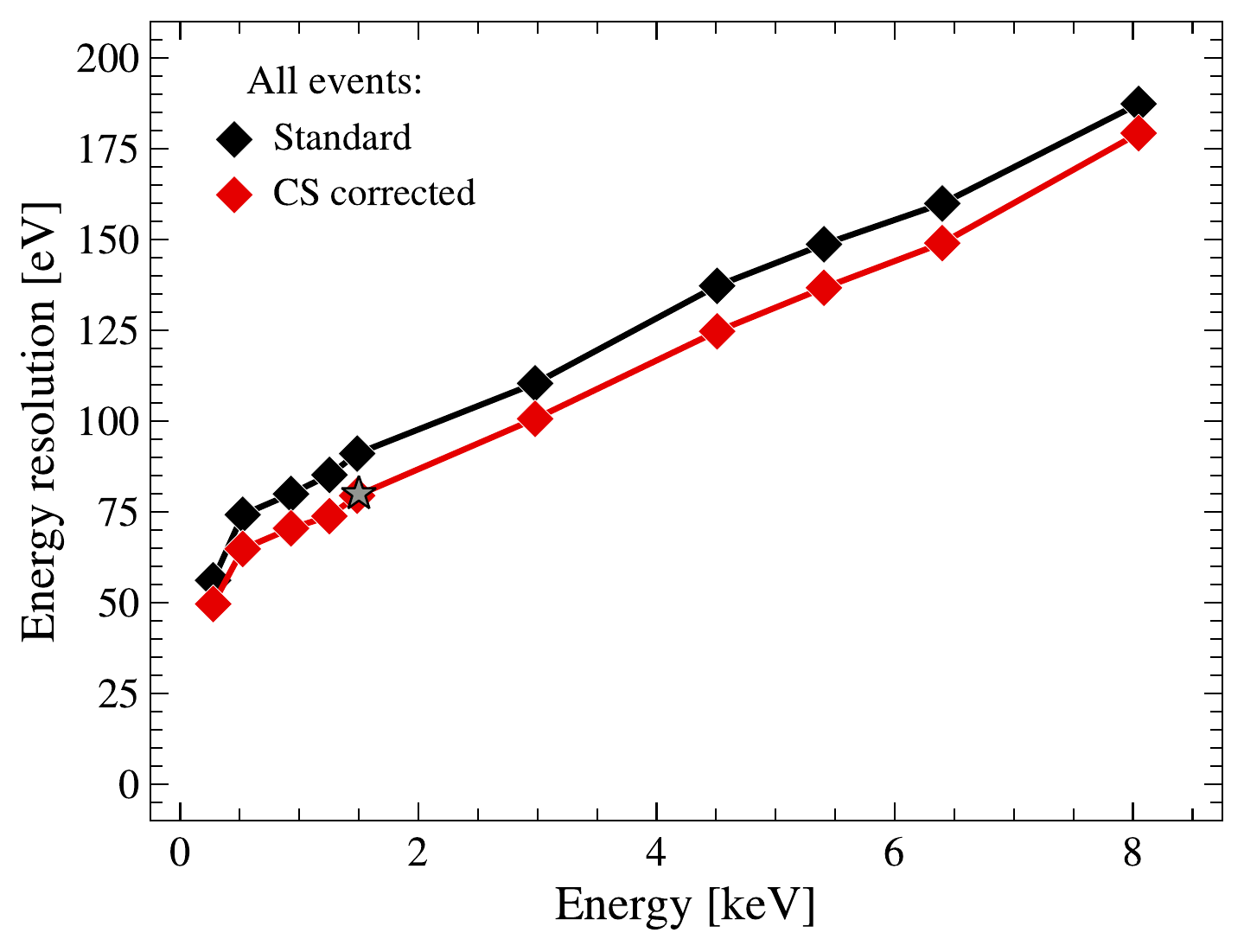}
        \end{subfigure}
        \caption{Energy resolution ($\Delta E$ in eV) as a function of energy. Circles (left panel) are for singles and diamonds (right panel) for all events (singles, doubles, triples and quadruples). Black and red colors represent the energy resolution without and with the CS correction. The grey star shows the instrument requirement of 80~eV at 1.5~keV.}
        \label{fig:fwhm_energy}
    \end{figure}

    The energy resolution ($\Delta E$) of the MXT camera was measured on ten intense X-ray lines from 0.28~keV to 8.05~keV produced by the Panter X-ray source. For each line, we determined the full width at half maximum (FWHM) with an optimized \texttt{Python} routine based on the \textit{lmfit} package \citep{newville2014a}.
    The routine fits a function composed of three to six parameters that correspond to a combination of a Gaussian function with a constant, linear or quadratic functions. The resulting best-fit models are visible in Fig.~\ref{fig:all_spec}. All fluorescence lines produced by the X-ray tube lie on a background signal (\textit{Bremsstrahlung} radiation) with an intensity depending on the source configuration (e.g., anode type, intensity or voltage, ...). To mitigate its influence on the spectral performance, we found that a model composed of 5 or 6 terms (i.e., a Gaussian plus a linear or quadratic function) minimized the $\chi^2$ statistic and provided the best results. 
    The Fig.~\ref{fig:fwhm_energy} shows for single and all events the MXT energy resolution as a function of the energy. For the single events, we found no significant effect of the CS correction on the spectral resolution compared to the standard calibration method. However, the benefit of this correction is more significant for the spectral performance of all events where, for all energy lines, the spectral resolution is considerably improved compared to the standard approach and closer to the performance of single events. The reason is that at first order, the CS correction shifts the energy line position of each multiplicity to its expected (theoretical) position, thus resulting in a combined spectrum with sharper lines and a better energy resolution.
    We determined an energy resolution of $\sim $73~eV and $\sim $79~eV for Al-K (1.49~keV) for single and all events, respectively. These performances are fully compliant with the instrument requirement of 80~eV at 1.5~keV and demonstrate the excellent spectral performance of MXT at beginning-of-life.
    It is worth noting that \cite{meidinger2006a} reported the state of the art in spectral performance for this generation (DUO) of pnCCD. They achieved a $\Delta E$ of 66~eV for single events and 74~eV for all events on the Al-K line at a similar temperature ($-70^{\circ}$C) to MXT but under more favorable conditions, i.e., not fully integrated in a space US components free designed instrument. \\ 
    Using all data sets (except the Cu-K) calibrated in energy and corrected from CS and CTI, we modeled the $\Delta E$ relation by considering the Fano noise and readout noise, i.e., with 2 free parameters ($F$, ENC) and a given pair creation energy ($\epsilon_w$) of 3.62~eV in silicon. For single events, we fit the following equation ($N_{\mathrm{split}}=1$):
    \begin{equation}
    \label{eq:deltaE}
        \Delta E_i= 2.35 \sqrt{\epsilon_w E_i F + N_{\mathrm{split}}(\epsilon_w \mathrm{ENC})^2} , \, \mathrm{with}\ E_i\ \mathrm{in\ eV}.
    \end{equation}
    The best-fit model returned a Fano factor $F$ of $0.131 \pm 0.003$, which is consistent with estimates reported in the literature \citep{kotov2018a} and an $\mathrm{ENC}$ of $4.9 \pm 0.2 \ \mathrm{e}^{-}_{\mathrm{rms}}$. This electronics noise includes the contribution of the detector leakage current (measurable with dark frames, Section.~\ref{sub:dark_noise_and_low_level_threshold}) and the calibration errors inherent to the methods and assumptions made (e.g., linear calibration).  \\
    Throughout the campaign, we also investigated the energy resolution as a function of detector temperature ($T_{\mathrm{det}}$). We acquired a series of measurements for three fluorescence lines (O-K, Cu-L and Al-K) and $T_{\mathrm{det}}=-75^{\circ}$C. We found very similar performance with no significant difference in the spectral performance between $-65^{\circ}$C and $-75^{\circ}$C which indicates that the electronics noise was minimal at the nominal temperature of the MXT ($-65^{\circ}$C) and was not the limiting factor of performances. However, this may not be the case at the end of the mission if the dark noise becomes more important due to radiation damage. \\
    Finally, the measurement on the C-K line confirmed the ability of MXT to detect photons down to 200~eV and the Cu-K$\beta$ line up to $\sim$9~keV. Both results demonstrate at the instrument level the low and high energy threshold of MXT and the compliance with the instrument requirements.

\section{Conclusions}
\label{sec:conclusions}
    During three weeks, we performed end-to-end tests to fully characterize the MXT instrument in its flight configuration and under space-like temperature and vacuum conditions at the MPE Panter facility. In this paper, we investigated the spectral performance of the MXT instrument with a series of measurements collected from 0.28~keV up to $\sim$9~keV for focused and defocused sources.
    First, we found that the pattern fractions for all multiplicities are in good agreement with the theoretical expectations obtained from MC simulations, confirming the high-quality of the collected data. We demonstrated the very good homogeneity and stability over time of the detector and its electronic chain regarding the dark noise and the calibration parameters.
    Then, we evaluated the spectral performance of MXT by optimizing the energy calibration process, especially by reducing the charge sharing effect induced by the LLT used to extract X-ray events. We verified the accuracy of our energy calibration process by measuring the positions of the calibrated lines with respect to their theoretical positions, and showed that the line positions are within the $\pm$20~eV instrument requirement up to $\sim$6~keV for single and all events. Above this energy, the trend of the line positions suggests a small non-linearity in the electronic readout chain. 
    Moreover, we found an energy resolution at 1.5~keV better than the MXT requirement of 80~eV for single and all events, and determined a CTI between $10^{-5}$--$10^{-4}$ for the current detector state. 
    Finally, we confirmed the ability of MXT to detect photons down to 200~eV and up to 10~keV. The end-to-end campaign has demonstrated the excellent spectral performance of MXT and its compliance with the instrument requirements, offering promising prospects for future science with MXT on GRBs and the time-domain astrophysics. \\
    The evolution of the spectral performance will be carefully tracked by spectral calibrations measurements during the mission. The 10 MeV-equivalent proton dose is expected to be $4\times10^{6}$ protons/cm\textsuperscript{2} after 5 years of mission. This will create displacement damage in the silicon detector causing dark current increase and charge trapping during the transfer. Both will increase the effective LLT which is essential for a fast source localization. Theoretical studies currently predict that the low-energy threshold of 200 eV can be maintained during the mission lifetime. A proton test campaign is in preparation to confirm this result experimentally.

\begin{acknowledgements}
    The authors thank all the people from CNES, CEA, MPE, IJCLab and the University of Leicester involved in the operations of the MXT instrument during the Panter calibration test campaign.
    We also thank O. Limousin and Y. Gutierrez for helpful discussions and advice about ECC.
    This research made use of Astropy\footnote{http://www.astropy.org}, a community-developed core \texttt{Python} package for Astronomy \citep{astropycollaboration2013a, astropycollaboration2018a}, of matplotlib, a \texttt{Python} library for publication quality graphics \citep{hunter2007a} and of NumPy \citep{harris2020a}.
\end{acknowledgements}

\section*{Declarations}
    \paragraph{Author’s Contribution: } 
    \label{par:author_s_contribution}
    All authors contributed to the study conception, design and data collection. Material preparation and analysis were performed by B.~Schneider, N.~Renault-Tinacci, D.~Götz, A.~Meuris and P.~Ferrando. The first draft of the manuscript was written by B.~Schneider and all authors commented on previous versions of the manuscript. All authors read and approved the final manuscript.
    \paragraph{Availability of Data and Materials: } 
    \label{par:availability_of_data_and_materials}
    All data analyzed during this study were collected at the MPE Panter facility and are part of the SVOM/MXT project.
    \paragraph{Funding: } 
    \label{par:funding}
    This work was funded by the French Space Agency (CNES), with the contribution of MPE, CEA, CNRS and the University of Leicester.

\bibliographystyle{aasjournal}
\bibliography{references}

\end{document}